\documentclass[%
aip,
jcp,%
amsmath,amssymb,
reprint%
,longbibliography,
floatfix]{revtex4-2}
\usepackage{graphicx} 
\usepackage{hyperref}
\hypersetup{colorlinks=true,
	linkcolor=magenta,
	allcolors=magenta}

\usepackage[outercaption]{sidecap} 
\sidecaptionvpos{figure}{c}
\usepackage{newtxtext}
\usepackage{newtxmath}
\usepackage[version=4]{mhchem}

\newcommand{\NPQtau}{$\mathrm{NPQ}_\tau$}
\newcommand{\Nanno}{\textit{N. oceanica}}
\usepackage{float}
\usepackage[caption = false]{subfig}

\begin{document}

\title{Kinetics of the xanthophyll cycle and its role in photoprotective memory and response}

\author{Audrey Short}
\thanks{These authors contributed equally: Audrey Short and Thomas P. Fay.}
\affiliation{Graduate Group in Biophysics, University of California, Berkeley, CA 94720 USA}
\affiliation{Molecular Biophysics and Integrated Bioimaging Division Lawrence Berkeley National Laboratory, Berkeley, CA 94720 USA}
\affiliation{Kavli Energy Nanoscience Institute, Berkeley, CA 94720 USA}
\author{Thomas P. Fay}
\thanks{These authors contributed equally: Audrey Short and Thomas P. Fay.}
\affiliation{Department of Chemistry, University of California Berkeley, CA 94720 USA}
\author{Thien Crisanto}
\affiliation{Molecular Biophysics and Integrated Bioimaging Division Lawrence Berkeley National Laboratory, Berkeley, CA 94720 USA}
\affiliation{Department of Plant and Microbial Biology, University of California, Berkeley, CA 94720 USA }
\affiliation{Howard Hughes Medical Institute, University of California, Berkeley, CA 94720 USA}
\author{Ratul Mangal}
\affiliation{Department of Chemistry, University of California Berkeley, CA 94720 USA}
\author{Krishna K. Niyogi}
\affiliation{Molecular Biophysics and Integrated Bioimaging Division Lawrence Berkeley National Laboratory, Berkeley, CA 94720 USA}
\affiliation{Department of Plant and Microbial Biology, University of California, Berkeley, CA 94720 USA }
\affiliation{Howard Hughes Medical Institute, University of California, Berkeley, CA 94720 USA}
\author{David T. Limmer}
\affiliation{Kavli Energy Nanoscience Institute, Berkeley, CA 94720 USA}
\affiliation{Department of Chemistry, University of California Berkeley, CA 94720 USA}
\affiliation{Chemical Science Division Lawrence Berkeley National Laboratory, Berkeley, CA 94720 USA}
\affiliation{Material Science Division Lawrence Berkeley National Laboratory, Berkeley, CA 94720 USA}
\author{Graham R. Fleming}
\email{grfleming@lbl.gov}
\affiliation{Graduate Group in Biophysics, University of California, Berkeley, CA 94720 USA}
\affiliation{Molecular Biophysics and Integrated Bioimaging Division Lawrence Berkeley National Laboratory, Berkeley, CA 94720 USA}
\affiliation{Kavli Energy Nanoscience Institute, Berkeley, CA 94720 USA}
\affiliation{Department of Chemistry, University of California Berkeley, CA 94720 USA}

\begin{abstract}

Efficiently balancing photochemistry and photoprotection is crucial for survival and productivity of photosynthetic organisms in the rapidly fluctuating light levels found in natural environments. The ability to respond quickly to sudden changes in light level is clearly advantageous. In the alga \textit{Nannochloropsis oceanica} we observed an ability to respond rapidly to sudden increases in light level which occur soon after a previous high-light exposure. This ability implies a kind of memory. In this work, we explore the xanthophyll cycle in \textit{N. oceanica} as a short-term photoprotective memory system. By combining snapshot fluorescence lifetime measurements with a biochemistry-based quantitative model, we show that short-term “memory” arises from the xanthophyll cycle. In addition, the model enables us to characterize the relative quenching abilities of the three xanthophyll cycle components. Given the ubiquity of the xanthophyll cycle in photosynthetic organisms the model described here will be of utility in improving our understanding of vascular plant and algal photoprotection with important implications for crop productivity.
\end{abstract}

\maketitle

\begin{SCfigure*}[0.4][t]
    \includegraphics[width=0.725\textwidth]{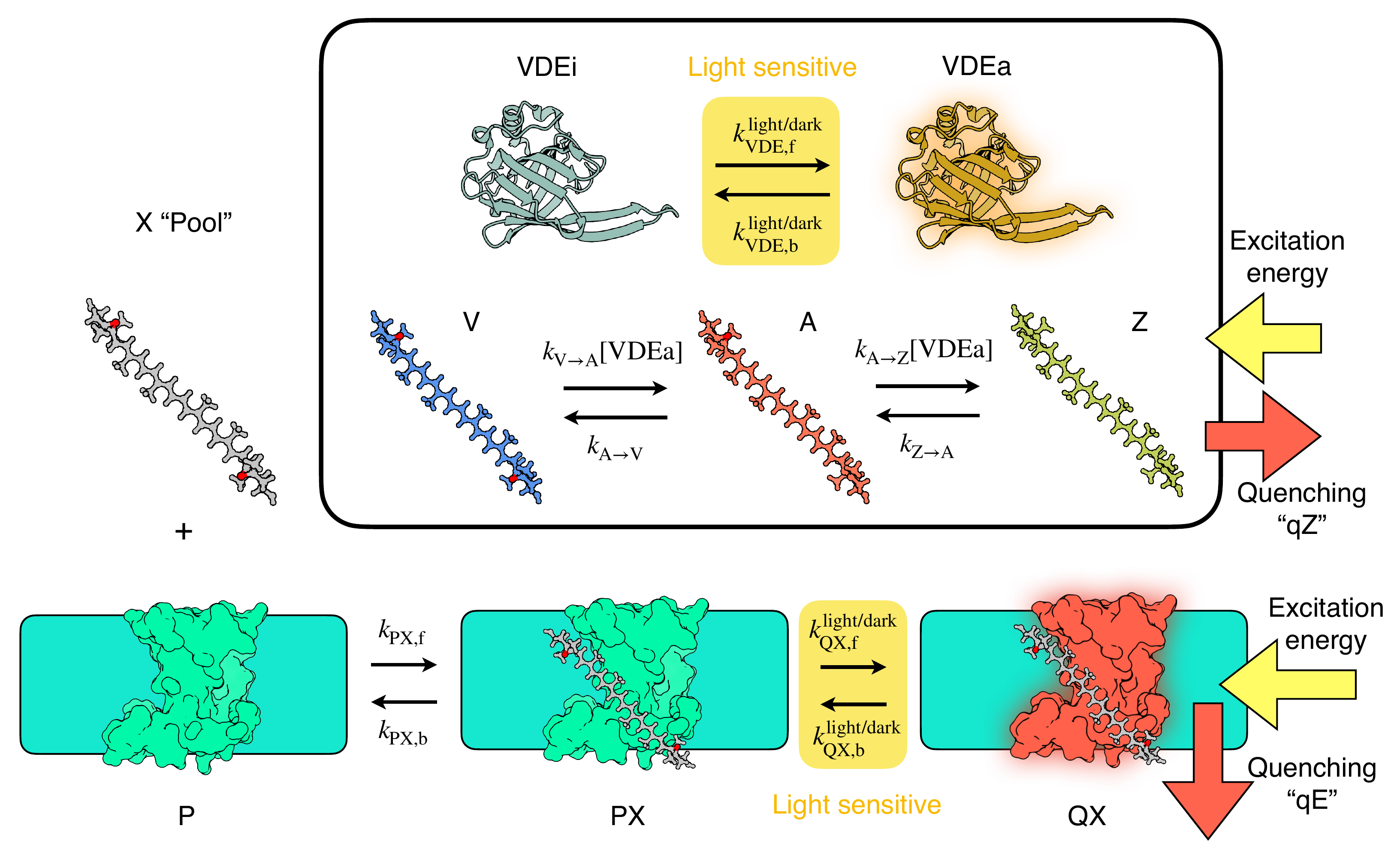}
    \caption{Illustration showing the processes included in the xanthophyll cycle based model. The xanthophyll (X) binds to the protein (P) reversibly to form a protein-xanthophyll complex (PX). In response to light this can convert into an active quencher form (QX). When not bound to the protein, the xanthophylls interconvert between violaxanthin (V), antheraxanthin (A) and zeaxanthin (Z). The activation of the VDE enzyme, which controls the \ce{V\to A\to Z} processes, is dependent on light conditions, which alter the ratio of the active VDE enzyme (VDEa) and its inactive from (VDEi). The light sensitive steps in the model are highlighted in yellow. The species responsible for quenching, the QX complexes in qE and pool Z in qZ, also indicated byc.}
    \label{fig-model-scheme}
\end{SCfigure*}

\section{Introduction}

In high-intensity light photosynthetic organisms are unable to utilize all available energy for photochemistry. In order to minimize the formation of damaging reactive oxygen species, the excess energy is dissipated as heat through non-photochemical quenching (NPQ) pathways\cite{demmig-adams2006,Ledford2005}. The eustigmatophyte alga \textit{Nannochloropsis oceanica} has a relatively simple NPQ system\cite{chukhutsina2017,park2019a} in comparison to vascular land plants. It consists of two main components: a pH-sensing protein, potentially LHCX1, and the xanthophyll cycle. 
The xanthophyll cycle in \Nanno\ is a shared feature with higher plants, but this alga lacks additional features like state transitions or pigments like lutein and chlorophyll-\emph{b}.\cite{litvin2016,Vieler2012,llansola-portoles2017} This simplistic nature makes \textit{N. oceanica} an ideal model organism for studying the essential components of NPQ. 

The xanthophyll cycle in \textit{N. oceanica} consists of the same de-epoxidation steps, from violaxanthin (V) to antheraxanthin (A) to zeaxanthin (Z), and reverse epoxidation steps, as seen in green algae and plants.\cite{Vieler2012,demmig-adams2020_zea} The enzyme violaxanthin de-epoxidase (VDE), located in the thylakoid lumen, converts V to A to Z upon protonation under high-light (HL) stress. Simultaneously, zeaxanthin epoxidase (ZEP), which is found in the stroma and thought to be constitutively active, reverses the VAZ cycle by epoxidizing Z to A to V\cite{Yamamoto1979,demmig-adams1993,Goss2006} (Fig.~\ref{fig-model-scheme}). It is now well-established that the VAZ cycle correlates with activation of energy-dependent quenching, qE, in both \textit{N. oceanica}\cite{chukhutsina2017,park2019a} and more complex organisms.\cite{Nilkens2010,Goss2015,Perin2023} The fast activating, pH-dependent quenching, qE, in \Nanno\ also depends on the protein LHCX1.\cite{park2019a} 
The mechanism of sensing changes in the thylakoid membrane pH-gradient and whether or not LHCX1 can bind pigments is still under investigation,\cite{buck2019,giovagnetti2021,park2019a,Taddei2018,Lacour2020,Buck2021} however the vital role of Z together with a pH-sensing protein in qE is widely achknowledged.\cite{demmig-adams2020_zea,Perin2023}  
The accumulation of A and Z has been observed to correlate with an increase in NPQ throughout a diurnal cycle in plants,\cite{demmig-adams1993,Goss2006} and 
it has been proposed that an additional, slower activating and slow deactivating Z-dependent quenching process also operates in the absence of a pH-gradient sensing protein, termed qZ.\cite{Nilkens2010,Goss2015} However the precise roles of the three xanthophylls and the kinetics of their interconversion in NPQ is not well understood, which is surprising given the prevalence of this widespread three-state photoprotective system in photosynthetic organisms.

In previous work,\cite{Short2022} we utilized a simplified kinetic model of the VAZ cycle that did not include the intermediate A to understand NPQ in \Nanno. Despite this simplification, the model gave useful insights into the time-scales of processes involved in NPQ activation, and it could quantitatively predict the quenching response, as well as qualitatively predict changes in V and Z concentrations, in response to a variety of regular and irregular light/dark illumination sequences. However, when exploring how the response changed when the dark period was progressively lengthened, it became clear that \textit{N. oceanica} has short-term “memory” of previous HL exposure which could not be captured by the simplified two-xanthophyll model. This type of memory of previous exposure to stressor events, wherein some organisms remain primed for an extended period to quickly respond to further stress, has been observed for other stressors such as in drought conditions.\cite{Sadhukhan2022} 
Various plant species, including \textit{Smilax australis}, \textit{Monstera deliciosa}, \textit{Vinca minor} and \textit{Vinca major}, have been shown to possess a long-term memory of growth light conditions, which is strongly species dependent. This long-term memory manifests in xanthophyll pool size and composition as well as maximum NPQ levels,\cite{Demmig-Adams2022,Demmig-Adams2020} an effect we also found evidence for previously in \Nanno.\cite{Short2022} It has also been shown that in phytoplankton and algae possessing a simpler two-state xanthophyll cycle, the xanthophylls can act as a long-term memory of growth light conditons.\cite{Polimene2012,Bidigare2014,Galindo2017} In this work we aim to explore the details of short-term photoprotective memory (operating on time-scales $\lesssim 1$ hour), complementing existing studies on connections between longer-term light exposure memory and the xanthophyll cycle. 

We hypothesise that in response to light stress, the VAZ cycle, and the kinetics of the different de-epoxidation/epoxidation steps, may act as a memory of previous HL exposure.\cite{Esteban2015} Specifically we propose that the presence of A in a system could keep plants and algae primed to respond to further HL stress, due to the slow rates of transforming A back to V. The role of the partially de-epoxidised xanthophyll A in photoprotection has been difficult to investigate directly, however work on plants has suggested that both A and Z correlate with NPQ in plants,\cite{Demmig-Adams2022,Demmig-Adams1996} but in this work we also aim to further elucidate its role in photoprotection. 
Previous work has shown the ratio of the rates from A$\to$Z to V$\to$A ranges from 4.5--6.3 times faster in various plant species, \cite{Hartel1996,siefermann1972,yamamoto1978} and the rate of epoxidation has been measured to be 1.4 times faster for Z than A.\cite{Goss2006} However precise measurements of these rates in \Nanno\ and their functional significance in NPQ and short-term memory of light stress have not been fully explored.

In this work, we aim to fully understand the role of xanthophyll cycle kinetics in photoprotective memory by considering the full VAZ cycle in modelling NPQ, and we show that differential rates of interconversion between the three xanthophylls are responsible for the multiple timescales of photoprotective memory. In a further step towards a comprehensive understanding of NPQ in \textit{N. oceanica}, the full VAZ model allows us to assess the relative quenching abilities of the three xanthophylls in the qE process, estimate the relative abundance of quenchers in the thylakoid membrane, and also quantify the relative contributions of LHCX1-dependent qE quenching and zeaxanthin-dependent qZ quenching in NPQ. In what follows, we start by briefly presenting our expanded model, then show how it accurately describes the HL stress responses of \Nanno, and how it encodes the functional role of the VAZ cycle in photoprotection. 

\section{Results}
\subsection{Kinetic model of xanthophyll-mediated photoprotection}

Motivated by measurements of xanthophyll concentrations and NPQ in response to light exposure (as presented in the next section), we have developed a new model for the coupled LHCX1-xanthophyll cycle photoprotection system in \Nanno, as is summarised schematically in Fig.~\ref{fig-model-scheme}. 
Before presenting any results, we briefly summarise the features of the model (details of the kinetic equations are given in the SI). 
In the predecessor to this model,\cite{Short2022} we neglected several important features that are included in the new model presented here, such as the intermediate A, which we will show plays an essential role in photoprotective memory, and the capability of each xanthophyll to act as a quencher, facilitated by LHCX1, which will be important for understanding the immediate response of \Nanno\ to light stress. Furthermore, we will show that the new model can quantitatively describe xanthophyll concentrations in cells, enabling us to estimate the absolute abundance of quenching sites in \Nanno\ and estimate its absolute quenching rate.

Overall the model involves 12 chemical species: the protein P, the three ``pool'' xanthophylls X = V, A and Z, three xanthophyll bound complexes PX in the non-quenching state and three in the quenching state QX, and the active (protonated) VDEa and inactive (unprotonated) VDEi forms of the VDE enzyme. Within the model, the protein P, binds the xanthophylls, X=V,A,Z, reversibly to form a complex PX. For simplicity we assume a single labile xanthophyll binding site per P, which we have found that this is sufficient to interpret the available experimental data. This PX complex is activated under HL (high light) conditions to reversibly form an active quencher, establishing the \ce{PX <=> QX} equilibrium, which we assume arises due to protonation and conformational changes. Previous work has identified LHCX1 as an essential component in activating the protein P, in the qE quenching mechanism,\cite{Short2022,park2019a,Goss2006} although the actual active quencher PX/QX could involve other proteins, especially since it is not known if LHCX1 binds pigments, and alternatively LHCX1 may just induce the conformational changes in P to activate quenching. Thus the precise identity of PX/QX is open to interpretation. The total fluorescence decay rate $\tau_\mathrm{F}(t)^{-1}$of chlorophylls in the membrane at a given time in the experiment $t$ is assumed to be related linearly to the concentration of the QX species, 
\begin{align}
\frac{1}{\tau_\mathrm{F}(t)} = \frac{1}{\tau_\mathrm{F,0}} \!+\! k_\mathrm{qE}([\ce{QV}](t)\!+\![\ce{QA}](t)\!+\![\ce{QZ}](t)) \!+\! k_{\mathrm{qZ}}[\ce{Z}](t), 
\end{align}
where $1/\tau_\mathrm{F,0}$ is the intrinsic fluorescence decay rate of chlorophyll (arising from both the dominant non-radiative and minor radiative pathways), and $k_\mathrm{qE}$ is the quenching rate constant for the QX complexes which mediate the LHCX1 and $\Delta$pH dependent qE quenching. We also incorporate zeaxanthin-dependent quenching, qZ, into the model by adding a quenching contribution which solely depends on the concentration of zeaxanthin in the ``pool''. The quenching rate constant for Z is denoted $k_{\mathrm{qZ}}$. We assume that qE and qZ mechanisms are non-radiative, dissipating chlorophyll excitation energy as heat into the environment. From this we can obtain the experimentally measured \NPQtau$=(\tau_\mathrm{F}(0) - \tau_\mathrm{F}(t)) / \tau_\mathrm{F}(t)$. We assume that whilst the extent to which PX converts to QX under HL conditions is dependent on X, the quenching rate of each complex in the chloroplast is the same. With the available \NPQtau\ data we found that it is not possible to ascertain whether the differences in total quenching capacity of the different QX species arise due to differences in quenching rate, or the positions of the \ce{PX <=> QX} equilibrium under HL conditions. Therefore, for simplicity we treat the quenching rate $k_\mathrm{qE}$ as being identical for all QX, and we also assume that the equilibrium constant for this process is zero in the dark. 

The interconversion of the xanthophylls is assumed to occur after unbinding of X from P, \ce{PX <=> P + X}. The X species in the model should be regarded as X in the pool on xanthophylls not bound to P. For example, X could be bound to other light-harvesting proteins from which it can unbind rapidly and reversibly. The xanthophylls in the pool can be de-epoxidised sequentially, from \ce{V \to A} and then \ce{A \to Z}, by VDEa, where the maximum turnover rate for the VDE enzyme is different for the two de-epoxidation steps. VDE is assumed to interconvert between VDEa and VDEi forms depending on light conditions. We model this as a simple two-state equilibrium with first-order rate laws for the activation and deactivation.  We also treat the epoxidation steps as sequential, first from \ce{Z \to A} then from \ce{A \to V}, and we assume that each epoxidation by the ZEP enzyme can be treated as a first-order rate process, with different epoxidation rates for Z and A. 

\subsection{Dynamical response of xanthophyll concentrations to light stress}
\begin{figure}
    \centering
    \includegraphics[width=0.475\textwidth]{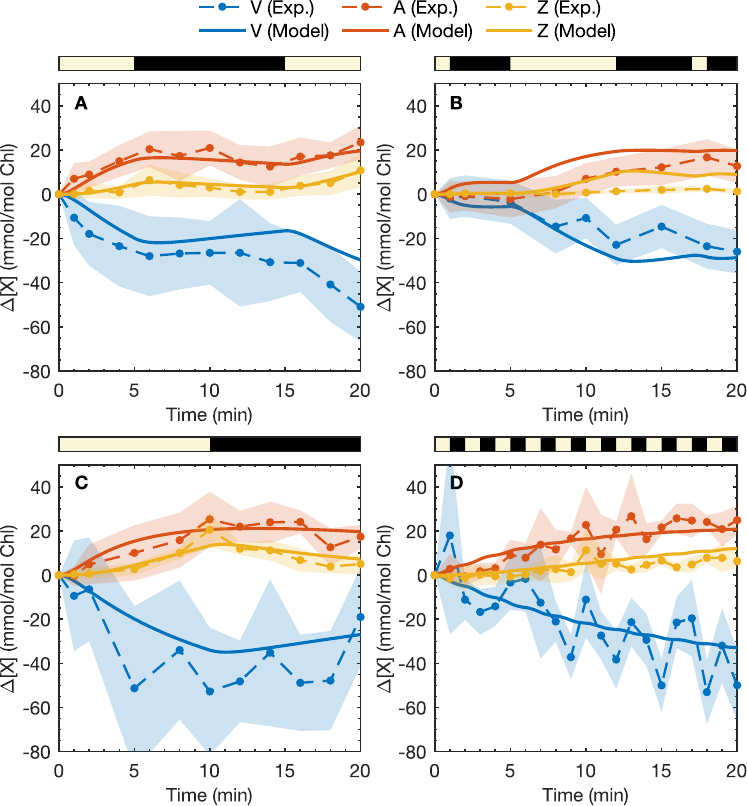}
    \caption{Experimental HPLC data for the change in xanthophyll concentration $\Delta[\ce{X}]$ as a function of time for four HL exposure sequences: A) 5 HL- 10 D- 5 HL, B) 1 HL- 4 D- 7 HL- 5 D- 1 HL- 2 D, C) 10 HL- 10 D, D) 1 HL-1 D (yellow shaded regions indicate the HL phases). Experimental results are shown as points and model predictions are shown as solid lines. Predictions correspond to the total xanthophyll cocnentrations, $\Delta[\ce{X}]_\mathrm{tot} = \Delta[\ce{X}] + \Delta[\ce{PX}] + \Delta[\ce{QX}]$. Experimental error bars (shaded regions) correspond to two standard errors of the mean. RMSD (root mean square deviations) in the fits are A) $\mathrm{RMSD}_{\mathrm{V}} = 11.2$, $\mathrm{RMSD}_{\mathrm{A}} = 8.6$, $\mathrm{RMSD}_{\mathrm{Z}} = 11.5$ B)
$\mathrm{RMSD}_{\mathrm{V}} = 6.8$, $\mathrm{RMSD}_{\mathrm{A}} = 11.5$, $\mathrm{RMSD}_{\mathrm{Z}} = 3.1$ C)
$\mathrm{RMSD}_{\mathrm{V}} = 14.6$, $\mathrm{RMSD}_{\mathrm{A}} = 9.5$, $\mathrm{RMSD}_{\mathrm{Z}} = 8.4$ and D)
$\mathrm{RMSD}_{\mathrm{V}} = 11.7$, $\mathrm{RMSD}_{\mathrm{A}} = 11.6$, $\mathrm{RMSD}_{\mathrm{Z}} = 10.7$ all in mmol/ mol Chl \textit{a}.
}
    \label{fig-hplc}
\end{figure}
\begin{table}[t]
    \centering
    \begin{tabular}{ccc}
        \hline
       Rate constant (min${}^{-1}$)   & HL conditions & Dark conditions
       \\
       \hline
       $k_{\ce{V}\to\ce{A},\mathrm{max}}$ & $0.092\pm 0.02$   & $(9.1\pm 6.2) \times 10 ^{-5}$ \\
       $k_{\ce{A}\to\ce{Z},\mathrm{max}}$ & $0.14\pm 0.05$ & $(1.4\pm 1.0) \times 10 ^{-4}$\\
       $k_{\ce{Z}\to\ce{A}}$ & $(8.5\pm 3.0)\times 10^{-2}$ &  $(8.5\pm 3.0)\times 10^{-2}$\\ 
       $k_{\ce{A}\to\ce{V}}$ & $(5.1\pm 2.8)\times 10^{-2}$ & $(5.1\pm 2.8)\times 10^{-2}$ \\
       $k_{\ce{VDE}}$ & 1.3$\pm 0.8$ & 1.0$\pm 0.75$\\
       \hline
    \end{tabular}
    { \footnotesize\caption{Rate constants for xanthophyll interconversion steps for the full VAZ model. $k_{\ce{X}\to\ce{X}',\mathrm{max}}$ is defined as $k_{\ce{X}\to\ce{X}'}[\ce{VDEa}]_{\mathrm{max}}^{\mathrm{light/dark}}$, where $[\ce{VDEa}]_{\mathrm{max}}^{\mathrm{light/dark}}$ is the maximum concentration of VDEa under light/dark conditions. $k_{\ce{VDE}} = k_{\ce{VDE,f}}+k_{\ce{VDE,b}}$ is the rate constant for activation/deactivation of VDE, such that in a light/dark phase [VDEa] changes according to $[\mathrm{VDEa}](t)-[\ce{VDEa}](t_0) = ([\mathrm{VDEa}]_\mathrm{max} - [\ce{VDEa}](t_0)) e^{-k_{\ce{VDE}}(t-t_0)}$. Errors given are two standard errors in the mean from bootstrapping. }
    \label{tab-vaz-rates}}
\end{table}
\begin{SCfigure*}[0.3][t]
    \centering
    \includegraphics[width=0.725\textwidth]{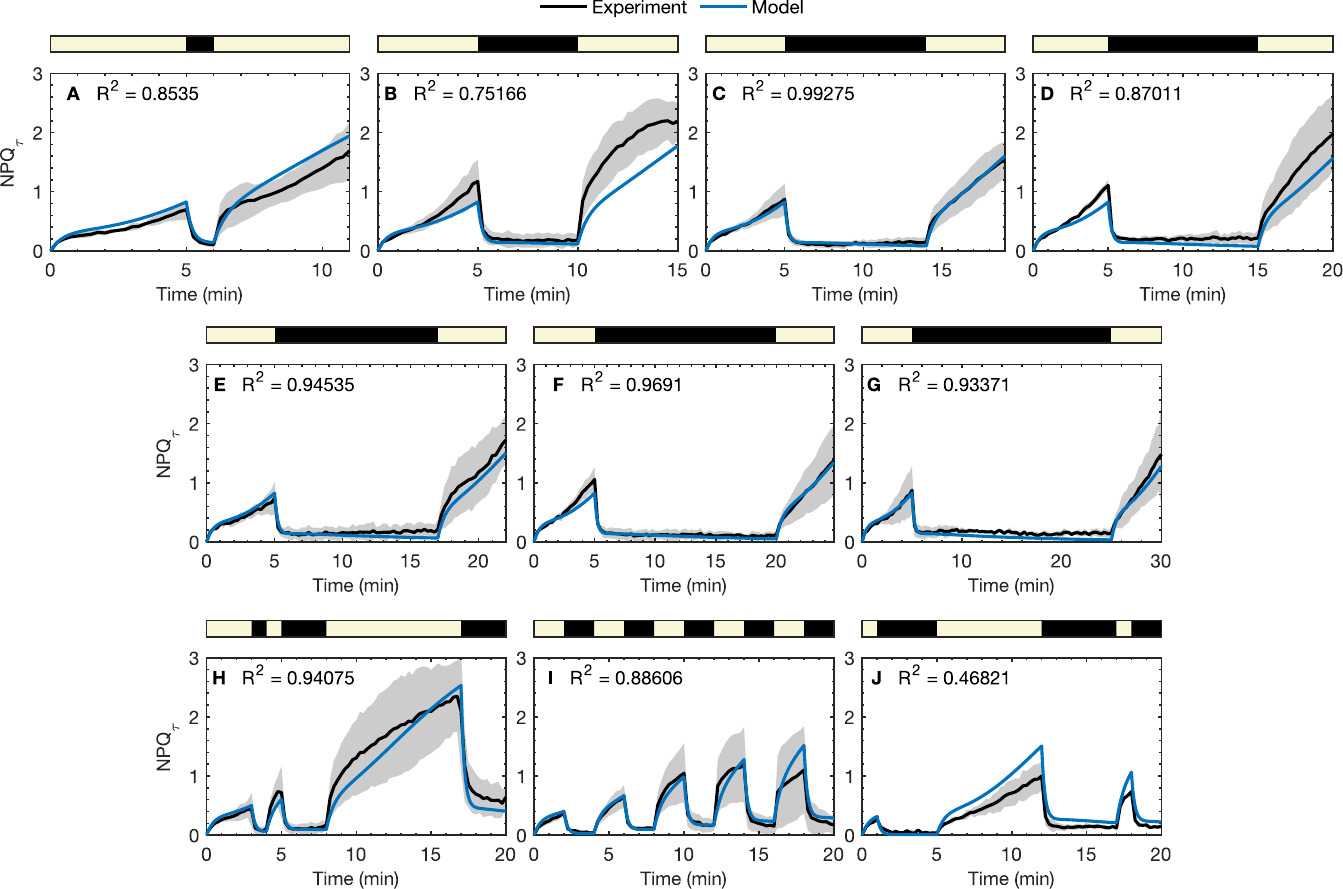}
    \caption[t]{Experimental \NPQtau\ data (black) together with the model predictions for the \NPQtau\ (blue) for various sequences of HL exposure/darkness for \Nanno. Yellow highlighted regions indicate HL phases of the experiments. A--G show data and model predictions for the 5 HL-$T$ D- 5HL sequences and H--J show three additional sequences, where HL denotes HL exposure and D denotes darkness, with number indicating the exposure time in min. RMSD values for the fits are A) 0.174 (n=3), B) 0.370 (n=3), C) 0.036 (n=3), D) 0.190 (n=3), E) 0.099 (n=3), F) 0.062 (n=3), G) 0.081 (n=3), H) 0.185 (n=3), I) 0.121 (n=3), J) 0.193 (n=3).}
    \label{fig-npq-seqs}
\end{SCfigure*}
In order to investigate the response of the xanthophyll cycle to fluctuating light conditions, we have measured the changes in concentrations of these pigments in \Nanno\ in response to four sequences of high-intensity light exposure: 5 HL- 10 D- 5 HL, 1 HL- 4 D- 7 HL- 5 D- 1 HL- 2 D, 10 HL- 10 D, and 1 HL-1 D, where HL denotes HL, D denotes darkness, and numbers indictate the duration of the exposure in minutes. The HPLC data showed a significant fraction of xanthophylls, particularly V, that remained constant over the time-scale of the experiment, which we believe corresponds to xanthophylls strongly bound to proteins other than LHCX1. The samples were dark-acclimated for 30 min prior to HL exposure to ensure minimal inital amounts of A and Z. Figure \ref{fig-hplc} shows the change in VAZ cycle carotenoids relative to their initial dark-acclimated values  (at $t=0$), i.e. $\Delta[\ce{X}] = [\ce{X}](t)-[\ce{X}](0)$ and $[\ce{X}](t)$ is the concentration of X at $t$. The experimental data show that $\Delta[\ce{A}]$ was greater than $\Delta$[Z] during HL exposures; $\Delta$[A] remained relatively constant during dark periods (Fig.~\ref{fig-hplc}), which shows a more rapid dynamical response to reduction in light exposure. In the 5 HL- 10 D- 5 HL sequence (Fig.~\ref{fig-hplc}.A), during the 10-minute dark period $\Delta$[Z] decreased almost entirely back to its dark-acclimated value whilst $\Delta$[A] remained constant for the first five minutes of darkness before it began to diminish. Both $\Delta$[A] and $\Delta$[Z] increased in response to the second HL exposure, and the rate of Z accumulation was greater than during the first HL period. Similarly in the 10 HL-10 D sequence (Fig.~\ref{fig-hplc}.C), $\Delta$[A] remained at a constant level compared to $\Delta$[Z], which decreased more rapidly back to its dark-acclimated concentration. In the 5 HL- 10 D - 5 HL and 10 HL - 10 D sequences, there was a small amount of continued accumulation of A and Z in the first dark phase for $\sim\! 1$ min, indicating a delayed deactivation of the de-epoxidation process, as we found previously in modelling the \NPQtau\ response of \Nanno.\cite{Short2022}

Rate constants for xanthophyll interconversion in the model were parameterised based on a reduced form of the full model, fitted to the experimental HPLC data, as detailed in the supporting information. The full model predictions for the HPLC data are also shown in Fig.~\ref{fig-hplc}, where we see the model mostly predicts the HPLC data within the experimental fluctuations, although in the 1 HL- 4 D- 7 HL- 5 D- 1 HL- 2 D sequence the model slightly overestimates $\Delta[\ce{A}]$ and $\Delta[\ce{Z}]$ after 1 min of light exposure (it should be noted that the fluctuations in xanthophyll concentrations in Fig.~\ref{fig-hplc} D do not correlate with the periodicity of light exposure on close inspection). 
In Table \ref{tab-vaz-rates} we summarise the maximum rates for the de-epoxidation processes, defined as $k_{\ce{X}\to\ce{X}'}[\ce{VDEa}]_{\mathrm{max}}^{\mathrm{light/dark}}$, and the epoxidation rates in the light and dark phases, and the rate constant for activation/deactivation (i.e. formation of VDEa from VDEi). We see that VDE activity increases by a factor of around 1000 in HL conditions, and that the VDE de-epoxidises A slightly faster than V, although the difference is small. Conversely for the epoxidation we see that Z is epoxidised nearly twice as fast as A. In our model we find that the VDE enzyme takes just over 1 min to activate and deactivate in both the light and dark phases, which is consistent with the continuing increase in A and Z concentrations observed at the start of the dark phases in the HPLC experiments.

\subsection{Modelling NPQ response of \Nanno\ to light exposure}\label{sec-res-npq}

Time correlated single photon counting (TCSPC) experiments were also performed on \Nanno\ to measure \NPQtau\ in response to sequences of HL/dark exposure. In addition to 20 minute regular and irregular light sequences that were utilized in previous work,\cite{Short2022} seven new HL/dark cycles were utilized to ascertain how long algae retain their ``photoprotective memory'' of previous HL exposure. The sequences had increasing dark durations ($T$ = 1, 5, 9, 10, 12, 15, 20 min) between two five-minute HL periods. The model was employed to describe \NPQtau\ dynamics of \Nanno\ in response to various HL/dark exposure sequences, with parameters determined by fitting a subset of the the \NPQtau\ sequences, namely the 5 HL-9 D-5 HL, 5 HL-15 D-5 HL, 3 HL-1 D-1 HL-3 D-9 HL-3 D, 1 HL-2 D-7 HL-5 D-1 HL-2 D, 2 HL-2 D sequences [Fig.~\ref{fig-npq-seqs} C,F,H,J]. Further details of this fitting procedure are given in the Methods section and supporting information. 

The experimental \NPQtau\ data are shown in Fig.~\ref{fig-npq-seqs}. We see rapid NPQ activation and deactivation in response to changes in light levels, occurring on a timescale of less than 1 min, together with a slower increase in \NPQtau\ during light exposure. The rapid component of \NPQtau\ activation and deactivation arising from the pH-sensing protein corresponds to the equilibration rate for the \ce{PX <=> QX} equilibrium in the model, given by $k_{\ce{QX}}^\mathrm{light/dark} = k_{\ce{QX},\mathrm{f}}^\mathrm{light/dark}+k_{\ce{QX},\mathrm{b}}^\mathrm{light/dark}$. This equilibration rate is 2.1 min${}^{-1}$ under light conditions and 4.7 min${}^{-1}$ in the dark which gives an activation time of $29$ s and a deactivation time of $13$ s. Experimental data for the 5 HL-$T$ D-5 HL sequences, Fig.~\ref{fig-npq-seqs}A--G, show how \NPQtau\ recovers after various dark durations, directly probing the photoprotective memory of \Nanno. 
The \NPQtau\ component recovered to its value at the end of the initial light period ($t= 5$ min) within 1 min upon secondary light exposure when dark durations were up to $T = 5$ min, and even with a 20 min dark duration the \NPQtau\ recovered within 3 min. 

In addition to the HPLC $\Delta[\ce{X}]_{\mathrm{tot}}$ data in Fig.~\ref{fig-hplc}, the model is able to predict the average \NPQtau\ levels for all the sequences as shown in Fig.~\ref{fig-npq-seqs}, including sequences other than those in the training datasets. 
Differences between the model predictions and experiments were generally comparable to the variability between experimental runs. For example at the end of the first five minutes of light exposure, \NPQtau\ in the 5 HL- $T$ D- 5HL sequences (Fig.~\ref{fig-npq-seqs}.A-G) the experimental \NPQtau varies between around 0.8 and 1.4, although some discrepancies may be attributed to shortcomings of the model. Specifically the over-prediction of \NPQtau\ for the 1 HL-2 D-7 HL-5 D-1 HL-2 D, 2 HL-2 D sequence [Fig.~\ref{fig-npq-seqs}.J] in the second light phase could be attributed to VDE activating too fast, as is seen in both the HPLC data and modelling [Fig.~\ref{fig-hplc}.B].

\begin{SCfigure*}[0.4][t]
    \includegraphics[width=0.6\textwidth]{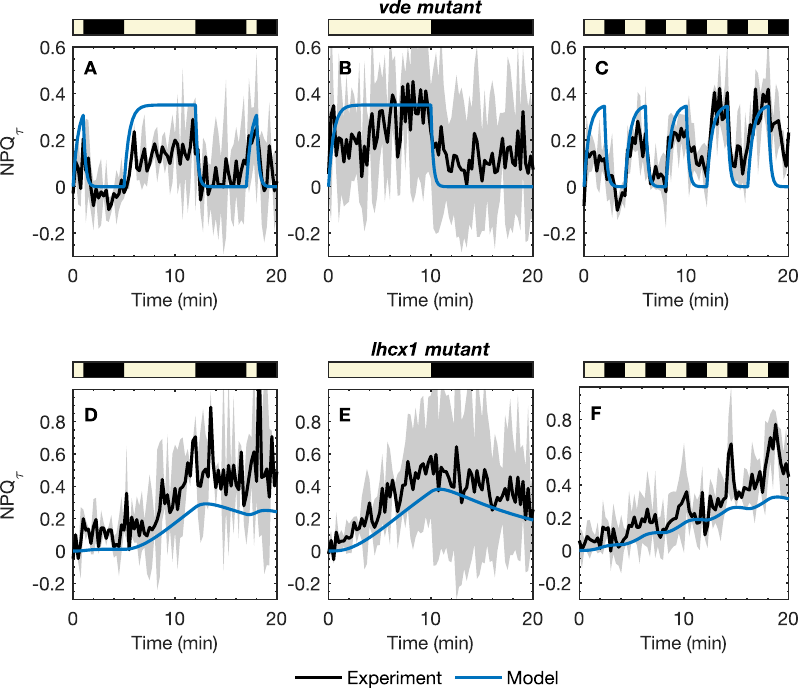}
    \caption{A--C) \NPQtau\ responses measured for the \emph{vde} mutant (black) together with model predictions (blue) for three sequences of light/dark exposure. Light/dark sequences: A) 1 HL-2 D-7 HL-5 D-1 HL-2 D (n=2), B) 10 HL-10 D (n=2), and C) 2 HL-2 D (n=3) $\times 5$. D--F) \NPQtau\ responses measured for the \emph{lhcx1} mutant (black) together with model predictions (blue) for three sequences of light/dark exposure. Light/dark sequences: A) 1 HL-2 D-7 HL-5 D-1 HL-2 D (n=2), B) 10 HL-10 D (n=3), and C) 2 HL-2 (n=3) D $\times 5$. RMSD for the model predictions are A) 0.134, B) 0.136, C) 0.118, D) 0.227, E) 0.139, F) 0.142.}
    \label{fig-npq-vde}
\end{SCfigure*}
\begin{table}[b]
    \centering
    \begin{tabular}{cccc}
        \hline 
      X  & Violaxanthin & Antheraxanthin & Zeaxanthin \\
       \hline
       $q_{\mathrm{X}}$ & $0.10\pm 0.02$ & $0.28\pm 0.08$  & $0.92\pm 0.05$ \\
       \hline
    \end{tabular}
    \caption{
    Quenching capacity, $q_{\ce{X}}$, for each of the xanthophylls. Errors given are two standard errors in the mean. }
    \label{tab-qX}
\end{table}
In the model the position of the \ce{PX <=> QX} equilibrium under HL conditions determines how well each of the xanthophylls can act as a quencher in qE. The maximum fraction of P-bound X that can exist in the QX state under HL conditions, denoted $q_{\ce{X}}$, determines the quenching capacity of each xanthophyll within our model. This can be expressed as
\begin{align}\label{eq-qX}
    q_{\ce{X}} = \frac{K^{\mathrm{light}}_{\ce{QX}}}{1+K^{\mathrm{light}}_{\ce{QX}}}
\end{align}
where $K_{\ce{QX}}^{\mathrm{light}}$ is the equilibrium constant for the \ce{PX <=> QX} process under HL conditions determined from fitting the model to the experimental data. In Table \ref{tab-qX} we list these values for our model under light and dark conditions, obtained from fitting the model to the experimental \NPQtau\ curves. From the $q_{\ce{X}}$ values we find that A is approximately three times more effective a quencher than V, and Z is nearly 10 times more effective than V. From the model we can also quantify the relative contributions of qE and qZ to the overall quenching, by the ratio of $k_{\mathrm{qZ}}$ to $k_{\mathrm{qE}}$, which is found to be $k_{\mathrm{qZ}}/k_{\mathrm{qE}} = 0.026\pm 0.005$.  

\subsection{NPQ in \Nanno\ mutants}

To further test the model, we have modified the wild type (WT) \Nanno\ parameterised model to predict the NPQ response of two \Nanno\ mutants: the \emph{vde} and \emph{lhcx1} mutants. 
The \NPQtau\ response of the \emph{vde} mutant, which has VDE knocked out preventing the accumulation of Z,\cite{} was modeled utilizing parameters obtained from the WT model with $k_{\ce{V\to A}}$ and $k_{\ce{A \to Z}}$ to zero. The \NPQtau\ response was measured for three HL/D sequences, shown in Fig.~\ref{fig-npq-vde}.A--C together with model predictions. 
Even in the absence of A and Z, \NPQtau\ increases near-instantaneously to around 0.3 in response to light, demonstrating the relevant role of LHCX1 in the \emph{vde} mutant. However, because of V’s low quenching capacity, the \NPQtau\ response is significantly smaller than that seen in WT, and there is no steady increase of \NPQtau\ over the duration of the experiment, unlike in the WT \Nanno. The model captures the \NPQtau\ response of the \emph{vde} mutant remarkably well, despite not being parameterised with these data. 

We have also modelled the \NPQtau\ response of the \emph{lhcx1} mutant, in which LHCX1 is not expressed and only zeaxanthin-mediated qZ quenching operates. This was modelled by simply setting $[\ce{P}]_{\mathrm{tot}} = 0$, removing the qE quenching process, while holding the total xanthophyll concentration constant. The experimental \NPQtau\ data and model predictions are shown in Fig.~\ref{fig-npq-vde}.D--F, where we see the model accurately captures the slow rise of \NPQtau\ in the light phases, arising from the build-up of Z during light exposure, and the slower decay in the dark phases due to slow epoxidation of Z. The success of the model in predicting the NPQ response of the \emph{lhcx1} mutant strongly supports the interpretation of the kinetic model species ``P'' as involving or at least requiring LHCX1 to function.
\section{Discussion}

Our combined experimental and kinetic model results suggest that photoprotective memory in \Nanno\ can be explained qualitatively with a simple three-state model. The three-state system should consist of a poor quencher (V), a modest quencher (A), and a good quencher (Z). After a sample has sufficiently accumulated the good quencher, during brief dark/low-light periods, Z remains before being converted back to the modest quencher (A), acting as short-term memory. However, during extended dark durations, Z will be converted almost entirely to A. Whilst A is also converted back to V, the A$\to$V transition occurs at a slower rate such that during another HL exposure occurs, the Z pool can form more rapidly from the reservoir of A. 
We can also see this dynamic represented in the HPLC data (Fig.~\ref{fig-hplc}). By adding the intermediate step in the VAZ cycle, the model not only more accurately reflects the biochemical processes but also allows for the short-term photoprotective memory, over time-scales between 1 min to $\sim\!30$ min, to be modelled and understood. 


From our experiments and modeling, we have also been able to determine the relative quenching capacities of the different xanthophylls. We find that V facilitates a weak but rapid response to changes in HL. The \emph{vde} mutant demonstrates that even without an effective quencher like Z, there is still an \NPQtau\ response to fluctuating light. In very short bursts of HL, V may act as the main quencher but the switch between its roles in photochemistry and photoprotection is determined by the pH gradient, as suggested previously.\cite{Horton2005}

\begin{figure}[t]
    \centering
    \includegraphics[width=0.475\textwidth]{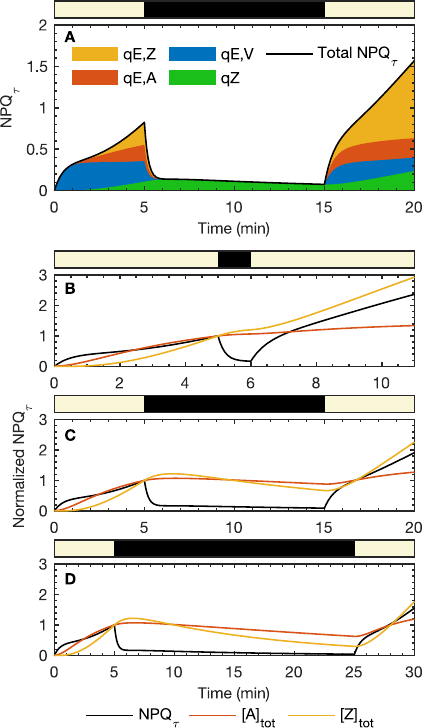}
    \caption{A) Contributions of each xanthophyll to the total \NPQtau\ as predicted by the model as a function of time for the 5 HL- 10D-5 HL sequence. B-D) \NPQtau, $[\ce{A}]_\mathrm{tot} = [\ce{A}] + [\ce{PA}] + [\ce{QA}]$, and $[\ce{Z}]_\mathrm{tot} = [\ce{Z}] + [\ce{PZ}] + [\ce{QZ}]$, predicted by the model for three 5 HL-$T$ D-5 HL sequences of light/dark exposure: B) $T=1$ min, C) $T = 10$ min and D) $T = 20$ min.}
    \label{fig-npq-rec-model}
\end{figure}

As the intermediate step in the VAZ cycle, A’s role as a potential quencher in qE is often overlooked. With a quenching capacity of around 30\%, it is 3.5 times less efficient than Z (95\%) at dissipating excess energy. However, it plays a crucial role in photoprotection in facilitating NPQ recovery after long dark durations. In Fig.~\ref{fig-npq-rec-model} we show a breakdown of the \NPQtau\ response predicted by the model for the 5 HL-10 D-5 HL sequence, where we see at short times the main quencher in qE is actually V complexed with LHCX1, with contributions from A emerging at $t = 1$ min and Z at $t = 2$ min. After light exposures of more than 2 min, Z functions as the primary quencher, with small, but not insignificant, contributions from V and A. Whilst LHCX1-dependent qE makes the largest contribution to \NPQtau, qZ also makes a small contribution, and within the model this is the sole contributor to the long-lived \NPQtau\ response in the dark. Even for very long-time light exposure, the model predicts that LHCX1-dependent qE dominates over qZ, with qZ making up only $\sim\!\! 23\%$ of the total \NPQtau\ in this limit, whilst the LHCX1-Zeaxanthin qE accounts for the majority ($\sim\!\! 75\%$) of the limiting \NPQtau. It should be noted that this limit is based on extrapolating the model to light exposure times beyond those which we have investigated, which may not be reliable, and we also expect the relative contributions of qE and qZ to depend strongly on species and growth conditions, as has been found in studies of plants.\cite{Demmig-Adams2022,Havaux1999,Demmig-Adams2012} We have not suggested a microscopic model for the qZ process, although in the SI, Sec. S.4, we show how a quenching process depending on some other zeaxanthin binding protein (or protein complex) P' would be consistent with our simple model. Zeaxanthin binding to some other protein could activate qZ by directly quenching excitation energy, potentially via charge transfer, or inducing conformational changes in the protein that promote other quenching mechanisms.\cite{Lapillo2020,Cupellini2020,Cignoni2021}

An essential element of the three-state photoprotective memory system observed in \Nanno\ is the kinetics of xanthophyll cycle, which together with the quenching capacities of the xanthophylls creates an effective photoprotective system. 
Upon the first exposure to light, NPQ activation is limited by moving through two steps before Z, the primary quencher, is accumulated, where VDE activation and the V$\to$A step (with a half life of $\sim\!7$ min) control the initial rate of NPQ activation.
Z may still function as a moderate quencher in the dark through qZ, so fast conversion of Z$\to$A by ZEP (half life $\sim 8$ min) in the dark is necessary to facilitate efficient photosynthesis under low light conditions.  The slower kinetics of A$\to$V in the dark (with a half life $\sim 20$ min) enables A to function as a buffer, facilitating rapid NPQ reactivation if light levels fluctuate again to damaging levels. The fast A$\to$Z conversion by VDE on light exposure (with a half-life of $\sim 4$ min) also plays an essential role in photoprotective memory by enabling the buffer of A to be rapidly converted to an active quencher. Previous work in plants found the rate of de-epoxidation of A to be about 4 times faster than that of V,\cite{Hartel1996,siefermann1972,yamamoto1978} which is a much larger difference compared to the de-epoxidation rates that we have found, with de-epoxidation of A being only about 1.5 times faster than that of V. However, VDE activity is influenced by the thylakoid lumen acidity, availability of ascorbate, and potentially unique species-specific differences, any of which could explain this discrepancy. Furthermore, because VDE is not active in the dark, the relative activity of ZEP on Z and A is far more relevant to photoprotective memory than the relative activity of VDE on V and A. On top of the slower time-scale kinetics of the VAZ cycle, which control the maximum quenching capacity of the system, very rapid responses to light fluctuations, on time-scales of around 1 min or less, are facilitated by protonation and subsequent conformational changes of the quenching protein which binds the xanthophylls.

From the model we can directly probe how the total A and Z concentrations vary during the 5 HL-$T$ D- 5HL sequences to demonstrate the functional role of xanthophyll cycle kinetics in photoprotective memory. Here we show in Fig.~\ref{fig-npq-rec-model} the model \NPQtau\, and the total A and Z concentrations normalized by their values at $t = 5$ min. For very short dark phase ($T = 1$ min, Fig.~\ref{fig-npq-rec-model}.B) Z continues to accumulate (due to the finite deactivation time of VDE in our model), acting as short-term light exposure memory and the \NPQtau\ recovers very rapidly upon re-illumination. For intermediate and longer lengths of dark duration ($T = 10$ min, Fig.~\ref{fig-npq-rec-model}.C and $T = 20$ min, Fig.~\ref{fig-npq-rec-model}.D), the quencher Z decreases but A remains steady, presumably acting as a buffer, and thus as a short-term memory for excess light exposure, and facilitating a fast response to HL in the second light phase. In these cases, that the \NPQtau\ response in the second HL phase correlates most strongly with the A concentration, and not the Z concentration. In the supporting information, Fig.~S2, we show the experimental and model \NPQtau\ recovery, averaged over the first minute of HL, in the second light phase for the 5 HL-$T$ D- 5HL sequences, as a function of dark duration $T$. From this we have extracted (see SI for details) an \NPQtau\ memory time-scale of $
\sim\!22$ min, which matches the model $\ce{A}\to\ce{V}$ time-scale given by $1/k_{\mathrm{A\to V}} = 19.9$ min. This strongly suggests that antheraxanthin acts as a short-term memory for light exposure, with the $\ce{A}\to\ce{V}$ step of the xanthophyll cycle controlling the effective memory time-scale. It has previously been observed that xanthophyll composition correlates with photoprotection, long- and medium-term light-exposure memory and light-levels during growth in plants\cite{Demmig-Adams2012,Demmig-Adams2020}, phytoplankton\cite{Polimene2012,Bidigare2014} and algae.\cite{Galindo2017} We can now however add to this picture that the kinetics of the xanthophyll cycle also plays an important role in short-term photoprotective memory. 

 

One important quantity we can estimate from this study is the lifetime of Chl-\emph{a} excitations on the active quenching complexes QX. Firstly from the HPLC data and model we obtain an estimate of the total concentration of P (possibly LHCX1 or LHCX1 in a complex with other proteins) in the system as approximately 0.6 mmol/mol Chl. Assuming roughly ten Chl-\emph{a} molecules per light-harvesting protein, this means the species P makes up approximately 1 in 30 light harvesting proteins in \Nanno. Using this ratio of P to the other light-harvesting proteins and assuming excitation energy diffusion between proteins is faster than quenching, we can estimate the lifetime of Chl-\emph{a}* on the active quenchers to be less than $\sim$10 ps (further details of this calculation are given in the SI, Sec.~S.4). This approximate time-scale is roughly consistent the quenching time-scale in HL acclimated \Nanno\ observed in transient-absorption experiments of $\sim$ 8 ps\cite{park2019a} (especially given the simplifying assumptions we use to deduce our estimate). Recent work has suggested that quenching can be limited by excitation energy redistribution within and between light-harvesting proteins,\cite{Bennett2018,Cupellini2020,Fay2022b} so the actual quenching process (likely either excitation energy transfer or charge transfer quenching\cite{park2019a}) may need to occur on an even shorter time-scale than this estimate. 

\section{Conclusion}

In this work, we have presented a model of xanthophyll cycle mediated non-photochemical quenching in \Nanno, which can both accurately describe the short and intermediate timescale \NPQtau\ responses of \Nanno\ to HL stress and the accompanying changes in xanthophyll concentrations. Employing a combination of experiments and modelling we have developing a deeper understanding of the photoprotective roles of the xanthophylls together with LHCX1. From this we have suggested a three-state model for short time-scale photoprotection in \Nanno, where the zeaxanthin-LHCX1 system acts as the primary quencher, with antheraxanthin acting as a short-term ``memory'' of HL stress capable of facilitating rapid response to fluctuations in light levels, and violaxanthin deactivating quenching under low-light conditions. This adds to the established picture of xanthophyll composition correlating for long-term memory of light-exposure.\cite{Demmig-Adams2022} Although we cannot conclusively identify the qE quencher, PX/QX, we can say that LHCX1 is an essential component of this system. We have also been able to estimate the chlorophyll excitation lifetime on active quenching proteins as less than $\sim$10 ps, as well as the relative abundance of quenchers in the thylakoid membrane. Evidence for zeaxanthin-dependent but LHCX1-independent qZ quenching has also been found, although its contribution to NPQ appears to be much smaller than that of LHCX1-dependent qE quenching. However, the propotion of qE or qZ contributions are going to vary depending on the species \cite{Demmig-Adams2023}. In order to implement a similar model of NPQ for use in vascular plants, more components need to be incorporated such as quenching due to lutein and state transitions,\cite{litvin2016,Vieler2012,llansola-portoles2017} which are not present in \Nanno. However, we believe the model presented here provides a basis for building a quantitative model of NPQ responses for plants and other photosynthetic organisms, which are mediated by the same xanthophyll cycle.


\subsection*{Acknowledgements}

The experimental work was supported by the U.S. Department of Energy, Office of Science, Chemical Sciences, Geosciences and Biosciences Division through FWP 449B to K.K.N and G.R.F. T.P.F. and D.T.L. were supported by the U.S. Department of Energy, Office of Science, Basic Energy Sciences, CPIMS Program Early Career Research Program under Award DE-FOA0002019. K.K.N. is an investigator of the Howard Hughes Medical Institute. We also thank Reviewer 1 for their useful comments in placing this work in broader context in the study of photoprotective memory.
		
\subsection*{Supporting Information}

Details of the model kinetic equations and simplification for analysis of HPLC data. Details of the model parameters. Analysis to obtain estimates of quencher abundance and quenching lifetime. Full HPLC data for the WT HL/dark sequences.

\subsection*{Data availability}

Source data are provided with this paper. All data presented in this manuscript is available at \url{www.doi.org/10.5281/zenodo.8284422}. 

\subsection*{Code availability}

All Matlab code used to run the model and produce figures in this manuscript is available at \url{www.doi.org/10.5281/zenodo.8284422}. 

\subsection*{Competing interests}

All authors declare no competing interests.

\subsection*{Author contributions}

GRF and AS conceived the research. AS performed all spectroscopic and HPLC experiments, and performed initial data analysis. TPF developed and implemented the model and performed final data analysis. TC prepared all algal samples and generated mutants. RM assisted AS in performing experiments. AS, TPF and GRF wrote the manuscript. AS, TPF, TC, GRF, DTL, and KKN discussed results and commented on the manuscript. DTL, KKN and GRF procured funding.

These authors contributed equally: Audrey Short, Thomas P. Fay.

\section*{Methods}
\subsection{Algal growth conditions}
\textit{N. oceanica} CCMP1779\cite{Vieler2012}  was obtained from the National Center for Marine Algae and Microbiota (https://ncma.bigelow.org/) and cultivated in F2N medium\cite{Kilian2011}. Liquid cultures were grown to 2-5 $\times$ 10${}^7$ cells/mL in continuous light at a photon flux density of 60 $\muup$mol photons m${}^{-2}$ s${}^{-1}$ at 22$^\circ$C or room temperature. 

The knock-out mutants \textit{vde} and \textit{lhcx1} were generated using homologous recombination of a hygromycin resistance cassette was used, with the addition of Cas9 RNP for \textit{lhcx1}. Further details of how the mutants were generated will be presented in a separate manuscript. 

\subsection{Time-correlated single photon counting}
Time-correlated single photon counting results in a histogram of Chl-\emph{a} fluorescence decay, which is then fit to a biexponential decay function yielding an average lifetime ($\bar{\tau}$). Fluorescence lifetime measurements were taken every 15 seconds to capture the change in fluorescence lifetimes as a function of HL exposure. The amplitude- weighted average lifetime of the Chl-\emph{a} fluorescence decay is converted into a unitless form, similar to that measured in the conventional pulse-amplitude modulation technique using the following equation: where $\bar{\tau}(0)$ and $\bar{\tau}(t)$ are the average lifetimes in the dark and at any time point \textit{t}, respectively, during the experiment.
\begin{align}\label{eq-NPQt}
    \mathrm{NPQ}_{\tau} = \frac{(\bar{\tau}(0)-\bar{\tau}(t))}{\bar{\tau}(t)}
\end{align}
An ultrafast Ti:sapphire coherent Mira 900 oscillator was pumped using a diode laser (Coherent Verdi G10, 532 nm). The center wavelength of the oscillator was 808 nm with a full width at half maximum of 9 nm. After frequency doubling the wavelength to 404 nm with a $\betaup$-barium borate crystal, the beam was split between the sample and a sync photodiode, which was used as a reference for snapshot measurements. Three synchronized shutters controlled the exposure of actinic light and the laser to the sample as well as to the microchannel plate-photomultiplier tube detector (Hamamatsu106 R3809U). The shutters were controlled by a LABVIEW software sequence. The detector was set to 680 nm to measure Chl-\emph{a} emission. During each snapshot, the laser and detection shutters were opened, allowing an excitation pulse with a power of 1.7 mW to saturate the reaction center for 1 second while the emission was recorded. During HL periods, samples were exposed to white light with an intensity of 885 $\muup$mol photons m${}^{-2}$ s${}^{-1}$ (Leica KL 1500 LCD, peak 648 nm, FWHM 220 nm ) by opening the actinic light shutter. 
The \textit{N. oceanica} sample was concentrated at ~ 40 $\muup$g Chl mL${}^{-1}$. To do this, 1 mL of \Nanno\ culture was pelleted for 5 minutes at room temperature at 14000 x RMP, flash frozen in liquid nitrogen, thawed at room temperature, and broken using FastPrep-24 (MP Biomedicals LLC) at 6.5 m/s for 60 seconds. The pellet was flash frozen and broken two more times. Chlorophyll was extracted from the broken cells using 1 mL of 80\% acetone, and total chlorophyll in the culture was quantified according to Porra \emph{et al.}.\cite{Porra1989} The culture was then concentrated by centrifuging for 5 minutes at room temperature at 4000 RPM. Samples were dark-acclimated for 30 minutes prior to the experiment and placed in the custom-built sample holder on a sample stage. The LABVIEW sequence was altered for each regular, irregular, and dark duration sequence run to control exposure to light fluctuations. For the \NPQtau\ experiments three technical replicates were performed for the WT and three for each mutant. Two experimental replicates were performed for the 5 HL-$T$ D-5 HL experiments and the training data for the model. Only one experimental replicate was performed for the mutants.
\subsection{High performance liquid chromatography}
Aliquots of \textit{N. oceanica} in F2N media were taken at various time points during several regular and irregular HL/dark duration actinic light sequences. Samples were then flash frozen in liquid nitrogen. After thawing, the samples were pelleted for 5 minutes at 4${}^\circ$C at 14000 $\times$ RPM to reach a cell count of $\sim$45-60$\times 10^6$. The cells were washed twice with \ce{dH2O} and pelleted at 14,000 $\times$ RPM for 5 minutes. The cells were again flash frozen and thawed at room temperature followed by breaking the cells using FastPrep-24 (MP Biomedicals LLC) at 6.5 m/s for 60 seconds. The bead beating step was repeated once before adding 200 $\muup$L of 100\% cold acetone. The samples were centrifuged for 10 minutes (14000 $\times$ RPM, 4${}^\circ$C), and the supernatant was filtered (0.2 $\muup$m nylon filter) into HPLC vials. The supernatant was separated on a Spherisorb S5 ODS1 4.6- $\times$ 250 mm cartridge column (Waters, Milford, MA) at 30${}^\circ$C. Analysis was completed using a modification of Garc{\'i}a-Plazaola and Becerril \cite{Garcia-Plazaola_Becerril} . Pigments were extracted with a linear gradient from 14\% solvent A (0.1M Tris-HCl pH 8.0 ddH20), 84\% (v/v) solvent B (acetonitrile), 2.0\% solvent C (methanol) for 15 minutes, to 68\% solvent C and 32\% solvent D (ethyl acetate) for 33 min, and then to 14\% solvent A (0.1M Tris-HCl pH 8.0 ddH2O), 84\% (v/v) solvent B (acetonitrile) , 2.0\% solvent C (methanol) for 19 min. The solvent flow rate was 1.2 mL min${}^{-1}$. Pigments were detected by A445 with reference at 550 nm by a diode array detector. Standard curves were prepared from concentrated pigments. The HPLC peaks were normalized to the total Chl-\emph{a} concentration.
\subsection{Model details}
Each step of the model given in Fig.~\ref{fig-model-scheme} is treated as an elementary reaction step in the 12 species model. As described on our previous work,\cite{Short2022} we cannot determine from these experiments the absolute concentration of VDE, so we replace the VDE species with a dynamical quantity $\alpha_{\ce{VDE}}(t)$ representing the activity of VDE at a time $t$ relative to its maximum possible value. We also work in a reduced unit system defined for species B by $[\widetilde{\ce{B}}] = \tau_{F}(0)k_{\ce{qE}}[\ce{B}]$, where $\tau_{F}(0)$ is the fluorescence lifetime at $t = 0$. With these reduced variables $\mathrm{NPQ}_\tau(t) = \Delta[\widetilde{\ce{QV}}](t) + \Delta[\widetilde{\ce{QA}}](t) + \Delta[\widetilde{\ce{QZ}}](t) + (k_{\ce{qZ}}/k_{\ce{qE}})\Delta[\widetilde{\ce{Z}}](t)$, where $\Delta[\widetilde{\ce{QX}}](t)$ is the change in reduced concentration of QX relative to the $t = 0$ value, and likewise for $\Delta[\widetilde{\ce{Z}}](t)$.

The model parameters were fitted by minimising the sum of square differences between the model \NPQtau\ and the experimental \NPQtau\ for the 5 HL-9 D-5 HL, 5 HL-15 D-5 HL, 3 HL-1 D-1 HL-3 D-9 HL-3 D, 1 HL-2 D-7 HL-5 D-1 HL-2 D, 2 HL-2 D sequences. Parameters for the VAZ interconversion steps were estimated from a least squares fit of a reduced model, which is a simple first order kinetic model with activation of the VDE enzyme, to the HPLC data (this is detailed in the SI). In the rest of the parameter fitting these parameters were constrained to only vary by 50\%. Additionally, to reduce the number of free parameters, the forward and backward binding rate constants $k_{\ce{PX,f}}$ and $k_{\ce{PX,b}}$, and the activation rate to form QX, $k_{\ce{QX}}^{\mathrm{light/dark}} = k_{\ce{QX,f}}^{\mathrm{light/dark}} + k_{\ce{QX,b}}^{\mathrm{light/dark}}$, were set to be independent of the species X. This way the equilibrium constant $K_{\ce{QX}}$ is the only parameter in the model controlling the quenching capacity of each xanthophyll. The remaining parameters were fitted first using Matlab's ``globalsearch'' function from an initial guess based on our previous model. This was then refined using the ``patternsearch'' algorithm. Errors in the fitted parameters were estimated by bootstrapping the experimental time series 1000 times. The conversion factor from reduced units to the mmol / mol Chl units the HPLC data are reported in was found using a simple least squares fit between the experimental HPLC and model $\Delta[\ce{X}]_{\mathrm{tot}}$ values shown in Fig.~\ref{fig-hplc}. Full details of the model kinetic equations and the full parameter set are given in the SI.

\bibliography{references.bib}

\begin{thebibliography}{43}%
\makeatletter
\providecommand \@ifxundefined [1]{%
 \@ifx{#1\undefined}
}%
\providecommand \@ifnum [1]{%
 \ifnum #1\expandafter \@firstoftwo
 \else \expandafter \@secondoftwo
 \fi
}%
\providecommand \@ifx [1]{%
 \ifx #1\expandafter \@firstoftwo
 \else \expandafter \@secondoftwo
 \fi
}%
\providecommand \natexlab [1]{#1}%
\providecommand \enquote  [1]{``#1''}%
\providecommand \bibnamefont  [1]{#1}%
\providecommand \bibfnamefont [1]{#1}%
\providecommand \citenamefont [1]{#1}%
\providecommand \href@noop [0]{\@secondoftwo}%
\providecommand \href [0]{\begingroup \@sanitize@url \@href}%
\providecommand \@href[1]{\@@startlink{#1}\@@href}%
\providecommand \@@href[1]{\endgroup#1\@@endlink}%
\providecommand \@sanitize@url [0]{\catcode `\\12\catcode `\$12\catcode
  `\&12\catcode `\#12\catcode `\^12\catcode `\_12\catcode `\%12\relax}%
\providecommand \@@startlink[1]{}%
\providecommand \@@endlink[0]{}%
\providecommand \url  [0]{\begingroup\@sanitize@url \@url }%
\providecommand \@url [1]{\endgroup\@href {#1}{\urlprefix }}%
\providecommand \urlprefix  [0]{URL }%
\providecommand \Eprint [0]{\href }%
\providecommand \doibase [0]{https://doi.org/}%
\providecommand \selectlanguage [0]{\@gobble}%
\providecommand \bibinfo  [0]{\@secondoftwo}%
\providecommand \bibfield  [0]{\@secondoftwo}%
\providecommand \translation [1]{[#1]}%
\providecommand \BibitemOpen [0]{}%
\providecommand \bibitemStop [0]{}%
\providecommand \bibitemNoStop [0]{.\EOS\space}%
\providecommand \EOS [0]{\spacefactor3000\relax}%
\providecommand \BibitemShut  [1]{\csname bibitem#1\endcsname}%
\let\auto@bib@innerbib\@empty
\bibitem [{\citenamefont {Demmig‐Adams}\ and\ \citenamefont
  {Adams}(2006)}]{demmig-adams2006}%
  \BibitemOpen
  \bibfield  {author} {\bibinfo {author} {\bibfnamefont {B.}~\bibnamefont
  {Demmig‐Adams}}\ and\ \bibinfo {author} {\bibfnamefont {W.~W.}\
  \bibnamefont {Adams}},\ }\bibfield  {title} {\enquote {\bibinfo {title}
  {{Photoprotection in an ecological context: the remarkable complexity of
  thermal energy dissipation}},}\ }\href
  {https://doi.org/10.1111/j.1469-8137.2006.01835.x} {\bibfield  {journal}
  {\bibinfo  {journal} {New Phytologist}\ }\textbf {\bibinfo {volume} {172}},\
  \bibinfo {pages} {11--21} (\bibinfo {year} {2006})}\BibitemShut {NoStop}%
\bibitem [{\citenamefont {Ledford}\ and\ \citenamefont
  {Niyogi}(2005)}]{Ledford2005}%
  \BibitemOpen
  \bibfield  {author} {\bibinfo {author} {\bibfnamefont {H.~K.}\ \bibnamefont
  {Ledford}}\ and\ \bibinfo {author} {\bibfnamefont {K.~K.}\ \bibnamefont
  {Niyogi}},\ }\bibfield  {title} {\enquote {\bibinfo {title} {{Singlet oxygen
  and photo‐oxidative stress management in plants and algae}},}\ }\href
  {https://doi.org/10.1111/j.1365-3040.2005.01374.x} {\bibfield  {journal}
  {\bibinfo  {journal} {Plant, Cell \& Environment}\ }\textbf {\bibinfo
  {volume} {28}},\ \bibinfo {pages} {1037--1045} (\bibinfo {year}
  {2005})}\BibitemShut {NoStop}%
\bibitem [{\citenamefont {Chukhutsina}\ \emph {et~al.}(2017)\citenamefont
  {Chukhutsina}, \citenamefont {Fristedt}, \citenamefont {Morosinotto},\ and\
  \citenamefont {Croce}}]{chukhutsina2017}%
  \BibitemOpen
  \bibfield  {author} {\bibinfo {author} {\bibfnamefont {V.~U.}\ \bibnamefont
  {Chukhutsina}}, \bibinfo {author} {\bibfnamefont {R.}~\bibnamefont
  {Fristedt}}, \bibinfo {author} {\bibfnamefont {T.}~\bibnamefont
  {Morosinotto}},\ and\ \bibinfo {author} {\bibfnamefont {R.}~\bibnamefont
  {Croce}},\ }\bibfield  {title} {\enquote {\bibinfo {title} {{Photoprotection
  strategies of the alga \textit{Nannochloropsis gaditana}}},}\ }\href
  {https://doi.org/10.1016/j.bbabio.2017.05.003} {\bibfield  {journal}
  {\bibinfo  {journal} {Biochimica et Biophysica Acta - Bioenergetics}\
  }\textbf {\bibinfo {volume} {1858}},\ \bibinfo {pages} {544---552} (\bibinfo
  {year} {2017})}\BibitemShut {NoStop}%
\bibitem [{\citenamefont {Park}\ \emph {et~al.}(2019)\citenamefont {Park},
  \citenamefont {Steen}, \citenamefont {Lyska}, \citenamefont {Fischer},
  \citenamefont {Endelman}, \citenamefont {Iwai}, \citenamefont {Niyogi},\ and\
  \citenamefont {Fleming}}]{park2019a}%
  \BibitemOpen
  \bibfield  {author} {\bibinfo {author} {\bibfnamefont {S.}~\bibnamefont
  {Park}}, \bibinfo {author} {\bibfnamefont {C.~J.}\ \bibnamefont {Steen}},
  \bibinfo {author} {\bibfnamefont {D.}~\bibnamefont {Lyska}}, \bibinfo
  {author} {\bibfnamefont {A.~L.}\ \bibnamefont {Fischer}}, \bibinfo {author}
  {\bibfnamefont {B.}~\bibnamefont {Endelman}}, \bibinfo {author}
  {\bibfnamefont {M.}~\bibnamefont {Iwai}}, \bibinfo {author} {\bibfnamefont
  {K.~K.}\ \bibnamefont {Niyogi}},\ and\ \bibinfo {author} {\bibfnamefont
  {G.~R.}\ \bibnamefont {Fleming}},\ }\bibfield  {title} {\enquote {\bibinfo
  {title} {{Chlorophyll–carotenoid excitation energy transfer and charge
  transfer in \textit{Nannochloropsis oceanica} for the regulation of
  photosynthesis}},}\ }\href {https://doi.org/10.1073/pnas.1819011116}
  {\bibfield  {journal} {\bibinfo  {journal} {Proceedings of the National
  Academy of Sciences of the United States of America}\ }\textbf {\bibinfo
  {volume} {116}},\ \bibinfo {pages} {3385---3390} (\bibinfo {year}
  {2019})}\BibitemShut {NoStop}%
\bibitem [{\citenamefont {Litvin}\ \emph {et~al.}(2016)\citenamefont {Litvin},
  \citenamefont {Bina}, \citenamefont {Herbstova},\ and\ \citenamefont
  {Gardian}}]{litvin2016}%
  \BibitemOpen
  \bibfield  {author} {\bibinfo {author} {\bibfnamefont {R.}~\bibnamefont
  {Litvin}}, \bibinfo {author} {\bibfnamefont {D.}~\bibnamefont {Bina}},
  \bibinfo {author} {\bibfnamefont {M.}~\bibnamefont {Herbstova}},\ and\
  \bibinfo {author} {\bibfnamefont {Z.}~\bibnamefont {Gardian}},\ }\bibfield
  {title} {\enquote {\bibinfo {title} {{Architecture of the light-harvesting
  apparatus of the eustigmatophyte alga \textit{Nannochloropsis oceanica}}},}\
  }\href {https://doi.org/10.1007/s11120-016-0234-1} {\bibfield  {journal}
  {\bibinfo  {journal} {Photosynthesis Research}\ }\textbf {\bibinfo {volume}
  {130}},\ \bibinfo {pages} {137---150} (\bibinfo {year} {2016})}\BibitemShut
  {NoStop}%
\bibitem [{\citenamefont {Vieler}\ \emph {et~al.}(2012)\citenamefont {Vieler},
  \citenamefont {Wu}, \citenamefont {Tsai}, \citenamefont {Bullard},
  \citenamefont {Cornish}, \citenamefont {Harvey}, \citenamefont {Reca},
  \citenamefont {Thornburg}, \citenamefont {Achawanantakun}, \citenamefont
  {Buehl}, \citenamefont {Campbell}, \citenamefont {Cavalier}, \citenamefont
  {Childs}, \citenamefont {Clark}, \citenamefont {Deshpande}, \citenamefont
  {Erickson}, \citenamefont {Ferguson}, \citenamefont {Handee}, \citenamefont
  {Kong}, \citenamefont {Li}, \citenamefont {Liu}, \citenamefont {Lundback},
  \citenamefont {Peng}, \citenamefont {Roston}, \citenamefont {Sanjaya},
  \citenamefont {Simpson}, \citenamefont {TerBush}, \citenamefont {Warakanont},
  \citenamefont {Zäuner}, \citenamefont {Farre}, \citenamefont {Hegg},
  \citenamefont {Jiang}, \citenamefont {Kuo}, \citenamefont {Lu}, \citenamefont
  {Niyogi}, \citenamefont {Ohlrogge}, \citenamefont {Osteryoung}, \citenamefont
  {Shachar-Hill}, \citenamefont {Sears}, \citenamefont {Sun}, \citenamefont
  {Takahashi}, \citenamefont {Yandell}, \citenamefont {Shiu},\ and\
  \citenamefont {Benning}}]{Vieler2012}%
  \BibitemOpen
  \bibfield  {author} {\bibinfo {author} {\bibfnamefont {A.}~\bibnamefont
  {Vieler}}, \bibinfo {author} {\bibfnamefont {G.}~\bibnamefont {Wu}}, \bibinfo
  {author} {\bibfnamefont {C.-H.}\ \bibnamefont {Tsai}}, \bibinfo {author}
  {\bibfnamefont {B.}~\bibnamefont {Bullard}}, \bibinfo {author} {\bibfnamefont
  {A.~J.}\ \bibnamefont {Cornish}}, \bibinfo {author} {\bibfnamefont
  {C.}~\bibnamefont {Harvey}}, \bibinfo {author} {\bibfnamefont {I.-B.}\
  \bibnamefont {Reca}}, \bibinfo {author} {\bibfnamefont {C.}~\bibnamefont
  {Thornburg}}, \bibinfo {author} {\bibfnamefont {R.}~\bibnamefont
  {Achawanantakun}}, \bibinfo {author} {\bibfnamefont {C.~J.}\ \bibnamefont
  {Buehl}}, \bibinfo {author} {\bibfnamefont {M.~S.}\ \bibnamefont {Campbell}},
  \bibinfo {author} {\bibfnamefont {D.}~\bibnamefont {Cavalier}}, \bibinfo
  {author} {\bibfnamefont {K.~L.}\ \bibnamefont {Childs}}, \bibinfo {author}
  {\bibfnamefont {T.~J.}\ \bibnamefont {Clark}}, \bibinfo {author}
  {\bibfnamefont {R.}~\bibnamefont {Deshpande}}, \bibinfo {author}
  {\bibfnamefont {E.}~\bibnamefont {Erickson}}, \bibinfo {author}
  {\bibfnamefont {A.~A.}\ \bibnamefont {Ferguson}}, \bibinfo {author}
  {\bibfnamefont {W.}~\bibnamefont {Handee}}, \bibinfo {author} {\bibfnamefont
  {Q.}~\bibnamefont {Kong}}, \bibinfo {author} {\bibfnamefont {X.}~\bibnamefont
  {Li}}, \bibinfo {author} {\bibfnamefont {B.}~\bibnamefont {Liu}}, \bibinfo
  {author} {\bibfnamefont {S.}~\bibnamefont {Lundback}}, \bibinfo {author}
  {\bibfnamefont {C.}~\bibnamefont {Peng}}, \bibinfo {author} {\bibfnamefont
  {R.~L.}\ \bibnamefont {Roston}}, \bibinfo {author} {\bibnamefont {Sanjaya}},
  \bibinfo {author} {\bibfnamefont {J.~P.}\ \bibnamefont {Simpson}}, \bibinfo
  {author} {\bibfnamefont {A.}~\bibnamefont {TerBush}}, \bibinfo {author}
  {\bibfnamefont {J.}~\bibnamefont {Warakanont}}, \bibinfo {author}
  {\bibfnamefont {S.}~\bibnamefont {Zäuner}}, \bibinfo {author} {\bibfnamefont
  {E.~M.}\ \bibnamefont {Farre}}, \bibinfo {author} {\bibfnamefont {E.~L.}\
  \bibnamefont {Hegg}}, \bibinfo {author} {\bibfnamefont {N.}~\bibnamefont
  {Jiang}}, \bibinfo {author} {\bibfnamefont {M.-H.}\ \bibnamefont {Kuo}},
  \bibinfo {author} {\bibfnamefont {Y.}~\bibnamefont {Lu}}, \bibinfo {author}
  {\bibfnamefont {K.~K.}\ \bibnamefont {Niyogi}}, \bibinfo {author}
  {\bibfnamefont {J.}~\bibnamefont {Ohlrogge}}, \bibinfo {author}
  {\bibfnamefont {K.~W.}\ \bibnamefont {Osteryoung}}, \bibinfo {author}
  {\bibfnamefont {Y.}~\bibnamefont {Shachar-Hill}}, \bibinfo {author}
  {\bibfnamefont {B.~B.}\ \bibnamefont {Sears}}, \bibinfo {author}
  {\bibfnamefont {Y.}~\bibnamefont {Sun}}, \bibinfo {author} {\bibfnamefont
  {H.}~\bibnamefont {Takahashi}}, \bibinfo {author} {\bibfnamefont
  {M.}~\bibnamefont {Yandell}}, \bibinfo {author} {\bibfnamefont {S.-H.}\
  \bibnamefont {Shiu}},\ and\ \bibinfo {author} {\bibfnamefont
  {C.}~\bibnamefont {Benning}},\ }\bibfield  {title} {\enquote {\bibinfo
  {title} {{Genome, Functional Gene Annotation, and Nuclear Transformation of
  the Heterokont Oleaginous Alga \textit{Nannochloropsis oceanica}
  CCMP1779}},}\ }\href {https://doi.org/10.1371/journal.pgen.1003064}
  {\bibfield  {journal} {\bibinfo  {journal} {PLoS Genetics}\ }\textbf
  {\bibinfo {volume} {8}},\ \bibinfo {pages} {e1003064} (\bibinfo {year}
  {2012})}\BibitemShut {NoStop}%
\bibitem [{\citenamefont {Llansola-Portoles}\ \emph {et~al.}(2017)\citenamefont
  {Llansola-Portoles}, \citenamefont {Litvin}, \citenamefont {Ilioaia},
  \citenamefont {Pascal}, \citenamefont {Bina},\ and\ \citenamefont
  {Robert}}]{llansola-portoles2017}%
  \BibitemOpen
  \bibfield  {author} {\bibinfo {author} {\bibfnamefont {M.~J.}\ \bibnamefont
  {Llansola-Portoles}}, \bibinfo {author} {\bibfnamefont {R.}~\bibnamefont
  {Litvin}}, \bibinfo {author} {\bibfnamefont {C.}~\bibnamefont {Ilioaia}},
  \bibinfo {author} {\bibfnamefont {A.~A.}\ \bibnamefont {Pascal}}, \bibinfo
  {author} {\bibfnamefont {D.}~\bibnamefont {Bina}},\ and\ \bibinfo {author}
  {\bibfnamefont {B.}~\bibnamefont {Robert}},\ }\bibfield  {title} {\enquote
  {\bibinfo {title} {{Pigment structure in the
  violaxanthin–chlorophyll-a-binding protein VCP}},}\ }\href
  {https://doi.org/10.1007/s11120-017-0407-6} {\bibfield  {journal} {\bibinfo
  {journal} {Photosynthesis Research}\ }\textbf {\bibinfo {volume} {134}},\
  \bibinfo {pages} {51--58} (\bibinfo {year} {2017})}\BibitemShut {NoStop}%
\bibitem [{\citenamefont {Demmig-Adams}\ \emph
  {et~al.}(2020{\natexlab{a}})\citenamefont {Demmig-Adams}, \citenamefont
  {Stewart}, \citenamefont {López-Pozo}, \citenamefont {Polutchko},\ and\
  \citenamefont {Adams}}]{demmig-adams2020_zea}%
  \BibitemOpen
  \bibfield  {author} {\bibinfo {author} {\bibfnamefont {B.}~\bibnamefont
  {Demmig-Adams}}, \bibinfo {author} {\bibfnamefont {J.~J.}\ \bibnamefont
  {Stewart}}, \bibinfo {author} {\bibfnamefont {M.}~\bibnamefont
  {López-Pozo}}, \bibinfo {author} {\bibfnamefont {S.~K.}\ \bibnamefont
  {Polutchko}},\ and\ \bibinfo {author} {\bibfnamefont {W.~W.}\ \bibnamefont
  {Adams}},\ }\bibfield  {title} {\enquote {\bibinfo {title} {{Zeaxanthin, a
  Molecule for Photoprotection in Many Different Environments}},}\ }\href
  {https://doi.org/10.3390/molecules25245825} {\bibfield  {journal} {\bibinfo
  {journal} {Molecules}\ }\textbf {\bibinfo {volume} {25}},\ \bibinfo {pages}
  {5825} (\bibinfo {year} {2020}{\natexlab{a}})}\BibitemShut {NoStop}%
\bibitem [{\citenamefont {Yamamoto}(1979)}]{Yamamoto1979}%
  \BibitemOpen
  \bibfield  {author} {\bibinfo {author} {\bibfnamefont {H.~Y.}\ \bibnamefont
  {Yamamoto}},\ }\bibfield  {title} {\enquote {\bibinfo {title} {{Biochemistry
  of the violaxanthin cycle in higher plants}},}\ }\href
  {https://doi.org/10.1016/b978-0-08-022359-9.50017-5} {\bibfield  {journal}
  {\bibinfo  {journal} {Pure and Applied Chemistry}\ ,\ \bibinfo {pages}
  {639--648}} (\bibinfo {year} {1979})}\BibitemShut {NoStop}%
\bibitem [{\citenamefont {Demmig-Adams}\ and\ \citenamefont
  {Adams}(1993)}]{demmig-adams1993}%
  \BibitemOpen
  \bibfield  {author} {\bibinfo {author} {\bibfnamefont {B.}~\bibnamefont
  {Demmig-Adams}}\ and\ \bibinfo {author} {\bibfnamefont {W.~W.}\ \bibnamefont
  {Adams}},\ }\bibfield  {title} {\enquote {\bibinfo {title} {{Carotenoids in
  Photosynthesis}},}\ }\href {https://doi.org/10.1007/978-94-011-2124-8\_7}
  {\bibfield  {journal} {\bibinfo  {journal} {Plant Physiol.}\ ,\ \bibinfo
  {pages} {206--251}} (\bibinfo {year} {1993})}\BibitemShut {NoStop}%
\bibitem [{\citenamefont {Goss}, \citenamefont {Lepetit},\ and\ \citenamefont
  {Wilhelm}(2006)}]{Goss2006}%
  \BibitemOpen
  \bibfield  {author} {\bibinfo {author} {\bibfnamefont {R.}~\bibnamefont
  {Goss}}, \bibinfo {author} {\bibfnamefont {B.}~\bibnamefont {Lepetit}},\ and\
  \bibinfo {author} {\bibfnamefont {C.}~\bibnamefont {Wilhelm}},\ }\bibfield
  {title} {\enquote {\bibinfo {title} {{Evidence for a rebinding of
  antheraxanthin to the light-harvesting complex during the epoxidation
  reaction of the violaxanthin cycle}},}\ }\href
  {https://doi.org/10.1016/j.jplph.2005.07.009} {\bibfield  {journal} {\bibinfo
   {journal} {Journal of Plant Physiology}\ }\textbf {\bibinfo {volume}
  {163}},\ \bibinfo {pages} {585--590} (\bibinfo {year} {2006})}\BibitemShut
  {NoStop}%
\bibitem [{\citenamefont {Nilkens}\ \emph {et~al.}(2010)\citenamefont
  {Nilkens}, \citenamefont {Kress}, \citenamefont {Lambrev}, \citenamefont
  {Miloslavina}, \citenamefont {Müller}, \citenamefont {Holzwarth},\ and\
  \citenamefont {Jahns}}]{Nilkens2010}%
  \BibitemOpen
  \bibfield  {author} {\bibinfo {author} {\bibfnamefont {M.}~\bibnamefont
  {Nilkens}}, \bibinfo {author} {\bibfnamefont {E.}~\bibnamefont {Kress}},
  \bibinfo {author} {\bibfnamefont {P.}~\bibnamefont {Lambrev}}, \bibinfo
  {author} {\bibfnamefont {Y.}~\bibnamefont {Miloslavina}}, \bibinfo {author}
  {\bibfnamefont {M.}~\bibnamefont {Müller}}, \bibinfo {author} {\bibfnamefont
  {A.~R.}\ \bibnamefont {Holzwarth}},\ and\ \bibinfo {author} {\bibfnamefont
  {P.}~\bibnamefont {Jahns}},\ }\bibfield  {title} {\enquote {\bibinfo {title}
  {Identification of a slowly inducible zeaxanthin-dependent component of
  non-photochemical quenching of chlorophyll fluorescence generated under
  steady-state conditions in \textit{Arabidopsis}},}\ }\href
  {https://doi.org/10.1016/j.bbabio.2010.01.001} {\bibfield  {journal}
  {\bibinfo  {journal} {Biochimica et Biophysica Acta (BBA) - Bioenergetics}\
  }\textbf {\bibinfo {volume} {1797}},\ \bibinfo {pages} {466--475} (\bibinfo
  {year} {2010})}\BibitemShut {NoStop}%
\bibitem [{\citenamefont {Goss}\ and\ \citenamefont
  {Lepetit}(2015)}]{Goss2015}%
  \BibitemOpen
  \bibfield  {author} {\bibinfo {author} {\bibfnamefont {R.}~\bibnamefont
  {Goss}}\ and\ \bibinfo {author} {\bibfnamefont {B.}~\bibnamefont {Lepetit}},\
  }\bibfield  {title} {\enquote {\bibinfo {title} {Biodiversity of {NPQ}},}\
  }\href {https://doi.org/10.1016/j.jplph.2014.03.004} {\bibfield  {journal}
  {\bibinfo  {journal} {Journal of Plant Physiology}\ }\textbf {\bibinfo
  {volume} {172}},\ \bibinfo {pages} {13--32} (\bibinfo {year}
  {2015})}\BibitemShut {NoStop}%
\bibitem [{\citenamefont {Perin}\ \emph {et~al.}(2023)\citenamefont {Perin},
  \citenamefont {Bellan}, \citenamefont {Michelberger}, \citenamefont {Lyska},
  \citenamefont {Wakao}, \citenamefont {Niyogi},\ and\ \citenamefont
  {Morosinotto}}]{Perin2023}%
  \BibitemOpen
  \bibfield  {author} {\bibinfo {author} {\bibfnamefont {G.}~\bibnamefont
  {Perin}}, \bibinfo {author} {\bibfnamefont {A.}~\bibnamefont {Bellan}},
  \bibinfo {author} {\bibfnamefont {T.}~\bibnamefont {Michelberger}}, \bibinfo
  {author} {\bibfnamefont {D.}~\bibnamefont {Lyska}}, \bibinfo {author}
  {\bibfnamefont {S.}~\bibnamefont {Wakao}}, \bibinfo {author} {\bibfnamefont
  {K.~K.}\ \bibnamefont {Niyogi}},\ and\ \bibinfo {author} {\bibfnamefont
  {T.}~\bibnamefont {Morosinotto}},\ }\bibfield  {title} {\enquote {\bibinfo
  {title} {Modulation of xanthophyll cycle impacts biomass productivity in the
  marine microalga \textit{Nannochloropsis}},}\ }\href
  {https://doi.org/10.1101/2022.08.16.504082} {\  (\bibinfo {year} {2023}),\
  10.1101/2022.08.16.504082},\ \bibinfo {note} {{Proceedings of the National
  Academy of Sciences}}\BibitemShut {NoStop}%
\bibitem [{\citenamefont {Buck}\ \emph {et~al.}(2019)\citenamefont {Buck},
  \citenamefont {Sherman}, \citenamefont {Bártulos}, \citenamefont {Serif},
  \citenamefont {Halder}, \citenamefont {Henkel}, \citenamefont {Falciatore},
  \citenamefont {Lavaud}, \citenamefont {Gorbunov}, \citenamefont {Kroth},
  \citenamefont {Falkowski},\ and\ \citenamefont {Lepetit}}]{buck2019}%
  \BibitemOpen
  \bibfield  {author} {\bibinfo {author} {\bibfnamefont {J.~M.}\ \bibnamefont
  {Buck}}, \bibinfo {author} {\bibfnamefont {J.}~\bibnamefont {Sherman}},
  \bibinfo {author} {\bibfnamefont {C.~R.}\ \bibnamefont {Bártulos}}, \bibinfo
  {author} {\bibfnamefont {M.}~\bibnamefont {Serif}}, \bibinfo {author}
  {\bibfnamefont {M.}~\bibnamefont {Halder}}, \bibinfo {author} {\bibfnamefont
  {J.}~\bibnamefont {Henkel}}, \bibinfo {author} {\bibfnamefont
  {A.}~\bibnamefont {Falciatore}}, \bibinfo {author} {\bibfnamefont
  {J.}~\bibnamefont {Lavaud}}, \bibinfo {author} {\bibfnamefont {M.~Y.}\
  \bibnamefont {Gorbunov}}, \bibinfo {author} {\bibfnamefont {P.~G.}\
  \bibnamefont {Kroth}}, \bibinfo {author} {\bibfnamefont {P.~G.}\ \bibnamefont
  {Falkowski}},\ and\ \bibinfo {author} {\bibfnamefont {B.}~\bibnamefont
  {Lepetit}},\ }\bibfield  {title} {\enquote {\bibinfo {title} {{Lhcx proteins
  provide photoprotection via thermal dissipation of absorbed light in the
  diatom \textit{Phaeodactylum tricornutum}}},}\ }\href
  {https://doi.org/10.1038/s41467-019-12043-6} {\bibfield  {journal} {\bibinfo
  {journal} {Nature Communications}\ }\textbf {\bibinfo {volume} {10}},\
  \bibinfo {pages} {4167} (\bibinfo {year} {2019})}\BibitemShut {NoStop}%
\bibitem [{\citenamefont {Giovagnetti}\ \emph {et~al.}(2021)\citenamefont
  {Giovagnetti}, \citenamefont {Jaubert}, \citenamefont {Shukla}, \citenamefont
  {Ungerer}, \citenamefont {Bouly}, \citenamefont {Falciatore},\ and\
  \citenamefont {Ruban}}]{giovagnetti2021}%
  \BibitemOpen
  \bibfield  {author} {\bibinfo {author} {\bibfnamefont {V.}~\bibnamefont
  {Giovagnetti}}, \bibinfo {author} {\bibfnamefont {M.}~\bibnamefont
  {Jaubert}}, \bibinfo {author} {\bibfnamefont {M.~K.}\ \bibnamefont {Shukla}},
  \bibinfo {author} {\bibfnamefont {P.}~\bibnamefont {Ungerer}}, \bibinfo
  {author} {\bibfnamefont {J.-P.}\ \bibnamefont {Bouly}}, \bibinfo {author}
  {\bibfnamefont {A.}~\bibnamefont {Falciatore}},\ and\ \bibinfo {author}
  {\bibfnamefont {A.~V.}\ \bibnamefont {Ruban}},\ }\bibfield  {title} {\enquote
  {\bibinfo {title} {{Biochemical and molecular properties of LHCX1, the
  essential regulator of dynamic photoprotection in diatoms}},}\ }\href
  {https://doi.org/10.1093/plphys/kiab425} {\bibfield  {journal} {\bibinfo
  {journal} {Plant Physiology}\ } (\bibinfo {year} {2021}),\
  10.1093/plphys/kiab425}\BibitemShut {NoStop}%
\bibitem [{\citenamefont {Taddei}\ \emph {et~al.}(2018)\citenamefont {Taddei},
  \citenamefont {Chukhutsina}, \citenamefont {Lepetit}, \citenamefont {Stella},
  \citenamefont {Bassi}, \citenamefont {van Amerongen}, \citenamefont {Bouly},
  \citenamefont {Jaubert}, \citenamefont {Finazzi},\ and\ \citenamefont
  {Falciatore}}]{Taddei2018}%
  \BibitemOpen
  \bibfield  {author} {\bibinfo {author} {\bibfnamefont {L.}~\bibnamefont
  {Taddei}}, \bibinfo {author} {\bibfnamefont {V.~U.}\ \bibnamefont
  {Chukhutsina}}, \bibinfo {author} {\bibfnamefont {B.}~\bibnamefont
  {Lepetit}}, \bibinfo {author} {\bibfnamefont {G.~R.}\ \bibnamefont {Stella}},
  \bibinfo {author} {\bibfnamefont {R.}~\bibnamefont {Bassi}}, \bibinfo
  {author} {\bibfnamefont {H.}~\bibnamefont {van Amerongen}}, \bibinfo {author}
  {\bibfnamefont {J.-P.}\ \bibnamefont {Bouly}}, \bibinfo {author}
  {\bibfnamefont {M.}~\bibnamefont {Jaubert}}, \bibinfo {author} {\bibfnamefont
  {G.}~\bibnamefont {Finazzi}},\ and\ \bibinfo {author} {\bibfnamefont
  {A.}~\bibnamefont {Falciatore}},\ }\bibfield  {title} {\enquote {\bibinfo
  {title} {Dynamic changes between two lhcx-related energy quenching sites
  control diatom photoacclimation},}\ }\href
  {https://doi.org/10.1104/pp.18.00448} {\bibfield  {journal} {\bibinfo
  {journal} {Plant Physiology}\ }\textbf {\bibinfo {volume} {177}},\ \bibinfo
  {pages} {953--965} (\bibinfo {year} {2018})}\BibitemShut {NoStop}%
\bibitem [{\citenamefont {Lacour}, \citenamefont {Babin},\ and\ \citenamefont
  {Lavaud}(2020)}]{Lacour2020}%
  \BibitemOpen
  \bibfield  {author} {\bibinfo {author} {\bibfnamefont {T.}~\bibnamefont
  {Lacour}}, \bibinfo {author} {\bibfnamefont {M.}~\bibnamefont {Babin}},\ and\
  \bibinfo {author} {\bibfnamefont {J.}~\bibnamefont {Lavaud}},\ }\bibfield
  {title} {\enquote {\bibinfo {title} {Diversity in xanthophyll cycle pigments
  content and related nonphotochemical quenching ({NPQ}) among microalgae:
  Implications for growth strategy and ecology},}\ }\href
  {https://doi.org/10.1111/jpy.12944} {\bibfield  {journal} {\bibinfo
  {journal} {Journal of Phycology}\ }\textbf {\bibinfo {volume} {56}},\
  \bibinfo {pages} {245--263} (\bibinfo {year} {2020})}\BibitemShut {NoStop}%
\bibitem [{\citenamefont {Buck}, \citenamefont {Kroth},\ and\ \citenamefont
  {Lepetit}(2021)}]{Buck2021}%
  \BibitemOpen
  \bibfield  {author} {\bibinfo {author} {\bibfnamefont {J.~M.}\ \bibnamefont
  {Buck}}, \bibinfo {author} {\bibfnamefont {P.~G.}\ \bibnamefont {Kroth}},\
  and\ \bibinfo {author} {\bibfnamefont {B.}~\bibnamefont {Lepetit}},\
  }\bibfield  {title} {\enquote {\bibinfo {title} {Identification of sequence
  motifs in lhcx proteins that confer {qE}-based photoprotection in the diatom
  \textit{Phaeodactylum tricornutum}},}\ }\href
  {https://doi.org/10.1111/tpj.15539} {\bibfield  {journal} {\bibinfo
  {journal} {Plant Journal}\ }\textbf {\bibinfo {volume} {108}},\ \bibinfo
  {pages} {1721--1734} (\bibinfo {year} {2021})}\BibitemShut {NoStop}%
\bibitem [{\citenamefont {Short}\ \emph {et~al.}(2022)\citenamefont {Short},
  \citenamefont {Fay}, \citenamefont {Crisanto}, \citenamefont {Hall},
  \citenamefont {Steen}, \citenamefont {Niyogi}, \citenamefont {Limmer},\ and\
  \citenamefont {Fleming}}]{Short2022}%
  \BibitemOpen
  \bibfield  {author} {\bibinfo {author} {\bibfnamefont {A.~H.}\ \bibnamefont
  {Short}}, \bibinfo {author} {\bibfnamefont {T.~P.}\ \bibnamefont {Fay}},
  \bibinfo {author} {\bibfnamefont {T.}~\bibnamefont {Crisanto}}, \bibinfo
  {author} {\bibfnamefont {J.}~\bibnamefont {Hall}}, \bibinfo {author}
  {\bibfnamefont {C.~J.}\ \bibnamefont {Steen}}, \bibinfo {author}
  {\bibfnamefont {K.~K.}\ \bibnamefont {Niyogi}}, \bibinfo {author}
  {\bibfnamefont {D.~T.}\ \bibnamefont {Limmer}},\ and\ \bibinfo {author}
  {\bibfnamefont {G.~R.}\ \bibnamefont {Fleming}},\ }\bibfield  {title}
  {\enquote {\bibinfo {title} {Xanthophyll-cycle based model of the rapid
  photoprotection of \textit{Nannochloropsis} in response to regular and
  irregular light/dark sequences},}\ }\href {https://doi.org/10.1063/5.0089335}
  {\bibfield  {journal} {\bibinfo  {journal} {The Journal of Chemical Physics}\
  }\textbf {\bibinfo {volume} {156}},\ \bibinfo {pages} {205102} (\bibinfo
  {year} {2022})}\BibitemShut {NoStop}%
\bibitem [{\citenamefont {Sadhukhan}\ \emph {et~al.}(2022)\citenamefont
  {Sadhukhan}, \citenamefont {Prasad}, \citenamefont {Mitra}, \citenamefont
  {Siddiqui}, \citenamefont {Sahoo}, \citenamefont {Kobayashi},\ and\
  \citenamefont {Koyama}}]{Sadhukhan2022}%
  \BibitemOpen
  \bibfield  {author} {\bibinfo {author} {\bibfnamefont {A.}~\bibnamefont
  {Sadhukhan}}, \bibinfo {author} {\bibfnamefont {S.~S.}\ \bibnamefont
  {Prasad}}, \bibinfo {author} {\bibfnamefont {J.}~\bibnamefont {Mitra}},
  \bibinfo {author} {\bibfnamefont {N.}~\bibnamefont {Siddiqui}}, \bibinfo
  {author} {\bibfnamefont {L.}~\bibnamefont {Sahoo}}, \bibinfo {author}
  {\bibfnamefont {Y.}~\bibnamefont {Kobayashi}},\ and\ \bibinfo {author}
  {\bibfnamefont {H.}~\bibnamefont {Koyama}},\ }\bibfield  {title} {\enquote
  {\bibinfo {title} {How do plants remember drought?}}\ }\href
  {https://doi.org/10.1007/s00425-022-03924-0} {\bibfield  {journal} {\bibinfo
  {journal} {Planta}\ }\textbf {\bibinfo {volume} {256}},\ \bibinfo {pages} {7}
  (\bibinfo {year} {2022})}\BibitemShut {NoStop}%
\bibitem [{\citenamefont {Demmig-Adams}\ \emph {et~al.}(2022)\citenamefont
  {Demmig-Adams}, \citenamefont {Polutchko}, \citenamefont {Stewart},\ and\
  \citenamefont {Adams}}]{Demmig-Adams2022}%
  \BibitemOpen
  \bibfield  {author} {\bibinfo {author} {\bibfnamefont {B.}~\bibnamefont
  {Demmig-Adams}}, \bibinfo {author} {\bibfnamefont {S.~K.}\ \bibnamefont
  {Polutchko}}, \bibinfo {author} {\bibfnamefont {J.~J.}\ \bibnamefont
  {Stewart}},\ and\ \bibinfo {author} {\bibfnamefont {W.~W.}\ \bibnamefont
  {Adams}},\ }\bibfield  {title} {\enquote {\bibinfo {title} {History of
  excess-light exposure modulates extent and kinetics of fast-acting
  non-photochemical energy dissipation},}\ }\href
  {https://doi.org/10.1007/s40502-022-00689-2} {\bibfield  {journal} {\bibinfo
  {journal} {Plant Physiology Reports}\ }\textbf {\bibinfo {volume} {27}},\
  \bibinfo {pages} {560--572} (\bibinfo {year} {2022})}\BibitemShut {NoStop}%
\bibitem [{\citenamefont {Demmig-Adams}\ \emph
  {et~al.}(2020{\natexlab{b}})\citenamefont {Demmig-Adams}, \citenamefont
  {Stewart}, \citenamefont {Adams}, \citenamefont {López-Pozo},\ and\
  \citenamefont {Polutchko}}]{Demmig-Adams2020}%
  \BibitemOpen
  \bibfield  {author} {\bibinfo {author} {\bibfnamefont {B.}~\bibnamefont
  {Demmig-Adams}}, \bibinfo {author} {\bibfnamefont {J.~J.}\ \bibnamefont
  {Stewart}}, \bibinfo {author} {\bibfnamefont {W.~W.}\ \bibnamefont {Adams}},
  \bibinfo {author} {\bibfnamefont {M.}~\bibnamefont {López-Pozo}},\ and\
  \bibinfo {author} {\bibfnamefont {S.~K.}\ \bibnamefont {Polutchko}},\
  }\bibfield  {title} {\enquote {\bibinfo {title} {Zeaxanthin, a molecule for
  photoprotection in many different environments},}\ }\href
  {https://doi.org/10.3390/MOLECULES25245825} {\bibfield  {journal} {\bibinfo
  {journal} {Molecules}\ }\textbf {\bibinfo {volume} {25}} (\bibinfo {year}
  {2020}{\natexlab{b}}),\ 10.3390/MOLECULES25245825}\BibitemShut {NoStop}%
\bibitem [{\citenamefont {Polimene}\ \emph {et~al.}(2012)\citenamefont
  {Polimene}, \citenamefont {Brunet}, \citenamefont {Allen}, \citenamefont
  {Butenschön}, \citenamefont {White},\ and\ \citenamefont
  {Llewellyn}}]{Polimene2012}%
  \BibitemOpen
  \bibfield  {author} {\bibinfo {author} {\bibfnamefont {L.}~\bibnamefont
  {Polimene}}, \bibinfo {author} {\bibfnamefont {C.}~\bibnamefont {Brunet}},
  \bibinfo {author} {\bibfnamefont {J.~I.}\ \bibnamefont {Allen}}, \bibinfo
  {author} {\bibfnamefont {M.}~\bibnamefont {Butenschön}}, \bibinfo {author}
  {\bibfnamefont {D.~A.}\ \bibnamefont {White}},\ and\ \bibinfo {author}
  {\bibfnamefont {C.~A.}\ \bibnamefont {Llewellyn}},\ }\bibfield  {title}
  {\enquote {\bibinfo {title} {Modelling xanthophyll photoprotective activity
  in phytoplankton},}\ }\href {https://doi.org/10.1093/plankt/fbr102}
  {\bibfield  {journal} {\bibinfo  {journal} {Journal of Plankton Research}\
  }\textbf {\bibinfo {volume} {34}},\ \bibinfo {pages} {196--207} (\bibinfo
  {year} {2012})}\BibitemShut {NoStop}%
\bibitem [{\citenamefont {Bidigare}\ \emph {et~al.}(2014)\citenamefont
  {Bidigare}, \citenamefont {Buttler}, \citenamefont {Christensen},
  \citenamefont {Barone}, \citenamefont {Karl},\ and\ \citenamefont
  {Wilson}}]{Bidigare2014}%
  \BibitemOpen
  \bibfield  {author} {\bibinfo {author} {\bibfnamefont {R.~R.}\ \bibnamefont
  {Bidigare}}, \bibinfo {author} {\bibfnamefont {F.~R.}\ \bibnamefont
  {Buttler}}, \bibinfo {author} {\bibfnamefont {S.~J.}\ \bibnamefont
  {Christensen}}, \bibinfo {author} {\bibfnamefont {B.}~\bibnamefont {Barone}},
  \bibinfo {author} {\bibfnamefont {D.~M.}\ \bibnamefont {Karl}},\ and\
  \bibinfo {author} {\bibfnamefont {S.~T.}\ \bibnamefont {Wilson}},\ }\bibfield
   {title} {\enquote {\bibinfo {title} {Evaluation of the utility of
  xanthophyll cycle pigment dynamics for assessing upper ocean mixing processes
  at station aloha},}\ }\href {https://doi.org/10.1093/plankt/fbu069}
  {\bibfield  {journal} {\bibinfo  {journal} {Journal of Plankton Research}\
  }\textbf {\bibinfo {volume} {36}},\ \bibinfo {pages} {1423--1433} (\bibinfo
  {year} {2014})}\BibitemShut {NoStop}%
\bibitem [{\citenamefont {Galindo}\ \emph {et~al.}(2017)\citenamefont
  {Galindo}, \citenamefont {Gosselin}, \citenamefont {Lavaud}, \citenamefont
  {Mundy}, \citenamefont {Else}, \citenamefont {Ehn}, \citenamefont {Babin},\
  and\ \citenamefont {Rysgaard}}]{Galindo2017}%
  \BibitemOpen
  \bibfield  {author} {\bibinfo {author} {\bibfnamefont {V.}~\bibnamefont
  {Galindo}}, \bibinfo {author} {\bibfnamefont {M.}~\bibnamefont {Gosselin}},
  \bibinfo {author} {\bibfnamefont {J.}~\bibnamefont {Lavaud}}, \bibinfo
  {author} {\bibfnamefont {C.~J.}\ \bibnamefont {Mundy}}, \bibinfo {author}
  {\bibfnamefont {B.}~\bibnamefont {Else}}, \bibinfo {author} {\bibfnamefont
  {J.}~\bibnamefont {Ehn}}, \bibinfo {author} {\bibfnamefont {M.}~\bibnamefont
  {Babin}},\ and\ \bibinfo {author} {\bibfnamefont {S.}~\bibnamefont
  {Rysgaard}},\ }\bibfield  {title} {\enquote {\bibinfo {title} {Pigment
  composition and photoprotection of arctic sea ice algae during spring},}\
  }\href {https://doi.org/10.3354/meps12398} {\bibfield  {journal} {\bibinfo
  {journal} {Marine Ecology Progress Series}\ }\textbf {\bibinfo {volume}
  {585}},\ \bibinfo {pages} {49--69} (\bibinfo {year} {2017})}\BibitemShut
  {NoStop}%
\bibitem [{\citenamefont {Esteban}\ \emph {et~al.}(2015)\citenamefont
  {Esteban}, \citenamefont {Moran}, \citenamefont {Becerril},\ and\
  \citenamefont {García-Plazaola}}]{Esteban2015}%
  \BibitemOpen
  \bibfield  {author} {\bibinfo {author} {\bibfnamefont {R.}~\bibnamefont
  {Esteban}}, \bibinfo {author} {\bibfnamefont {J.~F.}\ \bibnamefont {Moran}},
  \bibinfo {author} {\bibfnamefont {J.~M.}\ \bibnamefont {Becerril}},\ and\
  \bibinfo {author} {\bibfnamefont {J.~I.}\ \bibnamefont {García-Plazaola}},\
  }\bibfield  {title} {\enquote {\bibinfo {title} {{Versatility of carotenoids:
  An integrated view on diversity, evolution, functional roles and
  environmental interactions}},}\ }\href
  {https://doi.org/10.1016/j.envexpbot.2015.04.009} {\bibfield  {journal}
  {\bibinfo  {journal} {Environmental and Experimental Botany}\ }\textbf
  {\bibinfo {volume} {119}},\ \bibinfo {pages} {63--75} (\bibinfo {year}
  {2015})}\BibitemShut {NoStop}%
\bibitem [{\citenamefont {Demmig-Adams}\ and\ \citenamefont
  {Iii}(1996)}]{Demmig-Adams1996}%
  \BibitemOpen
  \bibfield  {author} {\bibinfo {author} {\bibfnamefont {B.}~\bibnamefont
  {Demmig-Adams}}\ and\ \bibinfo {author} {\bibfnamefont {W.~W.~A.}\
  \bibnamefont {Iii}},\ }\bibfield  {title} {\enquote {\bibinfo {title}
  {Xanthophyll cycle and light stress in nature: uniform response to excess
  direct sunlight among higher plant species},}\ }\href@noop {} {\bibfield
  {journal} {\bibinfo  {journal} {Planta}\ }\textbf {\bibinfo {volume} {198}},\
  \bibinfo {pages} {460--470} (\bibinfo {year} {1996})}\BibitemShut {NoStop}%
\bibitem [{\citenamefont {Hartel}\ \emph {et~al.}(1996)\citenamefont {Hartel},
  \citenamefont {Lokstein}, \citenamefont {Grimm},\ and\ \citenamefont
  {Rank}}]{Hartel1996}%
  \BibitemOpen
  \bibfield  {author} {\bibinfo {author} {\bibfnamefont {H.}~\bibnamefont
  {Hartel}}, \bibinfo {author} {\bibfnamefont {H.}~\bibnamefont {Lokstein}},
  \bibinfo {author} {\bibfnamefont {B.}~\bibnamefont {Grimm}},\ and\ \bibinfo
  {author} {\bibfnamefont {B.}~\bibnamefont {Rank}},\ }\bibfield  {title}
  {\enquote {\bibinfo {title} {{Kinetic Studies on the Xanthophyll Cycle in
  Barley Leaves (Influence of Antenna Size and Relations to Nonphotochemical
  Chlorophyll Fluorescence Quenching)}},}\ }\href
  {https://doi.org/10.1104/pp.110.2.471} {\bibfield  {journal} {\bibinfo
  {journal} {Plant Physiology}\ }\textbf {\bibinfo {volume} {110}},\ \bibinfo
  {pages} {471--482} (\bibinfo {year} {1996})}\BibitemShut {NoStop}%
\bibitem [{\citenamefont {Siefermann}(1972)}]{siefermann1972}%
  \BibitemOpen
  \bibfield  {author} {\bibinfo {author} {\bibfnamefont {D.}~\bibnamefont
  {Siefermann}},\ }\bibfield  {title} {\enquote {\bibinfo {title} {{Kinetic
  Studies on the Xanthophyll Cycle of Lemna gibba L. - Influence of
  Photosynthetic Oxygen and Supplied Reductor}},}\ }\href
  {https://doi.org/10.1007/978-94-010-2935-3\_65} {\bibfield  {journal}
  {\bibinfo  {journal} {Photosynthesis, two centuries after its discovery by
  Joseph Priestley}\ ,\ \bibinfo {pages} {629--635}} (\bibinfo {year}
  {1972})}\BibitemShut {NoStop}%
\bibitem [{\citenamefont {Yamamoto}\ and\ \citenamefont
  {Higashi}(1978)}]{yamamoto1978}%
  \BibitemOpen
  \bibfield  {author} {\bibinfo {author} {\bibfnamefont {H.}~\bibnamefont
  {Yamamoto}}\ and\ \bibinfo {author} {\bibfnamefont {R.}~\bibnamefont
  {Higashi}},\ }\bibfield  {title} {\enquote {\bibinfo {title} {{Violaxanthin
  de-epoxidase lipid composition and substrate specificity}},}\ }\href
  {https://doi.org/10.1016/0003-9861(78)90305-3} {\bibfield  {journal}
  {\bibinfo  {journal} {Archives of Biochemistry and Biophysics}\ }\textbf
  {\bibinfo {volume} {190}},\ \bibinfo {pages} {514--522} (\bibinfo {year}
  {1978})}\BibitemShut {NoStop}%
\bibitem [{\citenamefont {Horton}\ and\ \citenamefont
  {Ruban}(2005)}]{Horton2005}%
  \BibitemOpen
  \bibfield  {author} {\bibinfo {author} {\bibfnamefont {P.}~\bibnamefont
  {Horton}}\ and\ \bibinfo {author} {\bibfnamefont {A.}~\bibnamefont {Ruban}},\
  }\bibfield  {title} {\enquote {\bibinfo {title} {{Molecular design of the
  photosystem II light-harvesting antenna: photosynthesis and
  photoprotection}},}\ }\href {https://doi.org/10.1093/jxb/eri023} {\bibfield
  {journal} {\bibinfo  {journal} {Journal of Experimental Botany}\ }\textbf
  {\bibinfo {volume} {56}},\ \bibinfo {pages} {365--373} (\bibinfo {year}
  {2005})}\BibitemShut {NoStop}%
\bibitem [{\citenamefont {Havaux}\ \emph {et~al.}(1999)\citenamefont {Havaux},
  \citenamefont {†}, ,\ and\ \citenamefont {Niyogi}}]{Havaux1999}%
  \BibitemOpen
  \bibfield  {author} {\bibinfo {author} {\bibfnamefont {M.}~\bibnamefont
  {Havaux}}, \bibinfo {author} {\bibnamefont {†}}, ,\ and\ \bibinfo {author}
  {\bibfnamefont {K.~K.}\ \bibnamefont {Niyogi}},\ }\bibfield  {title}
  {\enquote {\bibinfo {title} {The violaxanthin cycle protects plants from
  photooxidative damage by more than one mechanism},}\ }\href {www.pnas.org.}
  {\bibfield  {journal} {\bibinfo  {journal} {Plant Biology}\ }\textbf
  {\bibinfo {volume} {96}},\ \bibinfo {pages} {8762--8767} (\bibinfo {year}
  {1999})}\BibitemShut {NoStop}%
\bibitem [{\citenamefont {Demmig-Adams}\ \emph {et~al.}(2012)\citenamefont
  {Demmig-Adams}, \citenamefont {Cohu}, \citenamefont {Muller},\ and\
  \citenamefont {Adams}}]{Demmig-Adams2012}%
  \BibitemOpen
  \bibfield  {author} {\bibinfo {author} {\bibfnamefont {B.}~\bibnamefont
  {Demmig-Adams}}, \bibinfo {author} {\bibfnamefont {C.~M.}\ \bibnamefont
  {Cohu}}, \bibinfo {author} {\bibfnamefont {O.}~\bibnamefont {Muller}},\ and\
  \bibinfo {author} {\bibfnamefont {W.~W.}\ \bibnamefont {Adams}},\ }\bibfield
  {title} {\enquote {\bibinfo {title} {Modulation of photosynthetic energy
  conversion efficiency in nature: From seconds to seasons},}\ }\href
  {https://doi.org/10.1007/s11120-012-9761-6} {\bibfield  {journal} {\bibinfo
  {journal} {Photosynthesis Research}\ }\textbf {\bibinfo {volume} {113}},\
  \bibinfo {pages} {75--88} (\bibinfo {year} {2012})}\BibitemShut {NoStop}%
\bibitem [{\citenamefont {Lapillo}\ \emph {et~al.}(2020)\citenamefont
  {Lapillo}, \citenamefont {Cignoni}, \citenamefont {Cupellini},\ and\
  \citenamefont {Mennucci}}]{Lapillo2020}%
  \BibitemOpen
  \bibfield  {author} {\bibinfo {author} {\bibfnamefont {M.}~\bibnamefont
  {Lapillo}}, \bibinfo {author} {\bibfnamefont {E.}~\bibnamefont {Cignoni}},
  \bibinfo {author} {\bibfnamefont {L.}~\bibnamefont {Cupellini}},\ and\
  \bibinfo {author} {\bibfnamefont {B.}~\bibnamefont {Mennucci}},\ }\bibfield
  {title} {\enquote {\bibinfo {title} {The energy transfer model of
  nonphotochemical quenching: Lessons from the minor cp29 antenna complex of
  plants},}\ }\href {https://doi.org/10.1016/j.bbabio.2020.148282} {\bibfield
  {journal} {\bibinfo  {journal} {Biochimica et Biophysica Acta (BBA) -
  Bioenergetics}\ }\textbf {\bibinfo {volume} {1861}},\ \bibinfo {pages}
  {148282} (\bibinfo {year} {2020})}\BibitemShut {NoStop}%
\bibitem [{\citenamefont {Cupellini}\ \emph {et~al.}(2020)\citenamefont
  {Cupellini}, \citenamefont {Calvani}, \citenamefont {Jacquemin},\ and\
  \citenamefont {Mennucci}}]{Cupellini2020}%
  \BibitemOpen
  \bibfield  {author} {\bibinfo {author} {\bibfnamefont {L.}~\bibnamefont
  {Cupellini}}, \bibinfo {author} {\bibfnamefont {D.}~\bibnamefont {Calvani}},
  \bibinfo {author} {\bibfnamefont {D.}~\bibnamefont {Jacquemin}},\ and\
  \bibinfo {author} {\bibfnamefont {B.}~\bibnamefont {Mennucci}},\ }\bibfield
  {title} {\enquote {\bibinfo {title} {Charge transfer from the carotenoid can
  quench chlorophyll excitation in antenna complexes of plants},}\ }\href
  {https://doi.org/10.1038/s41467-020-14488-6} {\bibfield  {journal} {\bibinfo
  {journal} {Nature Communications}\ }\textbf {\bibinfo {volume} {11}},\
  \bibinfo {pages} {662} (\bibinfo {year} {2020})}\BibitemShut {NoStop}%
\bibitem [{\citenamefont {Cignoni}\ \emph {et~al.}(2021)\citenamefont
  {Cignoni}, \citenamefont {Lapillo}, \citenamefont {Cupellini}, \citenamefont
  {Acosta-Gutiérrez}, \citenamefont {Gervasio},\ and\ \citenamefont
  {Mennucci}}]{Cignoni2021}%
  \BibitemOpen
  \bibfield  {author} {\bibinfo {author} {\bibfnamefont {E.}~\bibnamefont
  {Cignoni}}, \bibinfo {author} {\bibfnamefont {M.}~\bibnamefont {Lapillo}},
  \bibinfo {author} {\bibfnamefont {L.}~\bibnamefont {Cupellini}}, \bibinfo
  {author} {\bibfnamefont {S.}~\bibnamefont {Acosta-Gutiérrez}}, \bibinfo
  {author} {\bibfnamefont {F.~L.}\ \bibnamefont {Gervasio}},\ and\ \bibinfo
  {author} {\bibfnamefont {B.}~\bibnamefont {Mennucci}},\ }\bibfield  {title}
  {\enquote {\bibinfo {title} {A different perspective for nonphotochemical
  quenching in plant antenna complexes},}\ }\href
  {https://doi.org/10.1038/s41467-021-27526-8} {\bibfield  {journal} {\bibinfo
  {journal} {Nature Communications}\ }\textbf {\bibinfo {volume} {12}},\
  \bibinfo {pages} {7152} (\bibinfo {year} {2021})}\BibitemShut {NoStop}%
\bibitem [{\citenamefont {Bennett}, \citenamefont {Fleming},\ and\
  \citenamefont {Amarnath}(2018)}]{Bennett2018}%
  \BibitemOpen
  \bibfield  {author} {\bibinfo {author} {\bibfnamefont {D.~I.}\ \bibnamefont
  {Bennett}}, \bibinfo {author} {\bibfnamefont {G.~R.}\ \bibnamefont
  {Fleming}},\ and\ \bibinfo {author} {\bibfnamefont {K.}~\bibnamefont
  {Amarnath}},\ }\bibfield  {title} {\enquote {\bibinfo {title}
  {Energy-dependent quenching adjusts the excitation diffusion length to
  regulate photosynthetic light harvesting},}\ }\href
  {https://doi.org/10.1073/pnas.1806597115} {\bibfield  {journal} {\bibinfo
  {journal} {Proceedings of the National Academy of Sciences of the United
  States of America}\ }\textbf {\bibinfo {volume} {115}},\ \bibinfo {pages}
  {E9523--E9531} (\bibinfo {year} {2018})}\BibitemShut {NoStop}%
\bibitem [{\citenamefont {Fay}\ and\ \citenamefont {Limmer}(2022)}]{Fay2022b}%
  \BibitemOpen
  \bibfield  {author} {\bibinfo {author} {\bibfnamefont {T.~P.}\ \bibnamefont
  {Fay}}\ and\ \bibinfo {author} {\bibfnamefont {D.~T.}\ \bibnamefont
  {Limmer}},\ }\bibfield  {title} {\enquote {\bibinfo {title} {Coupled charge
  and energy transfer dynamics in light harvesting complexes from a hybrid
  hierarchical equations of motion approach},}\ }\href
  {https://doi.org/10.1063/5.0117659} {\bibfield  {journal} {\bibinfo
  {journal} {The Journal of Chemical Physics}\ }\textbf {\bibinfo {volume}
  {157}},\ \bibinfo {pages} {174104} (\bibinfo {year} {2022})}\BibitemShut
  {NoStop}%
\bibitem [{\citenamefont {López-Pozo}\ \emph {et~al.}(2023)\citenamefont
  {López-Pozo}, \citenamefont {Adams}, \citenamefont {Polutchko},\ and\
  \citenamefont {Demmig-Adams}}]{Demmig-Adams2023}%
  \BibitemOpen
  \bibfield  {author} {\bibinfo {author} {\bibfnamefont {M.}~\bibnamefont
  {López-Pozo}}, \bibinfo {author} {\bibfnamefont {W.~W.}\ \bibnamefont
  {Adams}}, \bibinfo {author} {\bibfnamefont {S.~K.}\ \bibnamefont
  {Polutchko}},\ and\ \bibinfo {author} {\bibfnamefont {B.}~\bibnamefont
  {Demmig-Adams}},\ }\bibfield  {title} {\enquote {\bibinfo {title}
  {{Terrestrial and Floating Aquatic Plants Differ in Acclimation to Light
  Environment}},}\ }\href {https://doi.org/10.3390/plants12101928} {\bibfield
  {journal} {\bibinfo  {journal} {Plants}\ }\textbf {\bibinfo {volume} {12}},\
  \bibinfo {pages} {1928} (\bibinfo {year} {2023})}\BibitemShut {NoStop}%
\bibitem [{\citenamefont {Kilian}\ \emph {et~al.}(2011)\citenamefont {Kilian},
  \citenamefont {Benemann}, \citenamefont {Niyogi},\ and\ \citenamefont
  {Vick}}]{Kilian2011}%
  \BibitemOpen
  \bibfield  {author} {\bibinfo {author} {\bibfnamefont {O.}~\bibnamefont
  {Kilian}}, \bibinfo {author} {\bibfnamefont {C.~S.~E.}\ \bibnamefont
  {Benemann}}, \bibinfo {author} {\bibfnamefont {K.~K.}\ \bibnamefont
  {Niyogi}},\ and\ \bibinfo {author} {\bibfnamefont {B.}~\bibnamefont {Vick}},\
  }\bibfield  {title} {\enquote {\bibinfo {title} {{High-efficiency homologous
  recombination in the oil-producing alga Nannochloropsis sp.}}}\ }\href
  {https://doi.org/10.1073/pnas.1105861108} {\bibfield  {journal} {\bibinfo
  {journal} {Proceedings of the National Academy of Sciences}\ }\textbf
  {\bibinfo {volume} {108}},\ \bibinfo {pages} {21265--21269} (\bibinfo {year}
  {2011})}\BibitemShut {NoStop}%
\bibitem [{\citenamefont {Porra}, \citenamefont {Thompson},\ and\ \citenamefont
  {Kriedemann}(1989)}]{Porra1989}%
  \BibitemOpen
  \bibfield  {author} {\bibinfo {author} {\bibfnamefont {R.}~\bibnamefont
  {Porra}}, \bibinfo {author} {\bibfnamefont {W.}~\bibnamefont {Thompson}},\
  and\ \bibinfo {author} {\bibfnamefont {P.}~\bibnamefont {Kriedemann}},\
  }\bibfield  {title} {\enquote {\bibinfo {title} {Determination of accurate
  extinction coefficients and simultaneous equations for assaying chlorophylls
  a and b extracted with four different solvents: verification of the
  concentration of chlorophyll standards by atomic absorption spectroscopy},}\
  }\href {https://doi.org/10.1016/S0005-2728(89)80347-0} {\bibfield  {journal}
  {\bibinfo  {journal} {Biochimica et Biophysica Acta (BBA) - Bioenergetics}\
  }\textbf {\bibinfo {volume} {975}},\ \bibinfo {pages} {384--394} (\bibinfo
  {year} {1989})}\BibitemShut {NoStop}%
\bibitem [{\citenamefont {García‐Plazaola}\ and\ \citenamefont
  {Becerril}(1999)}]{Garcia-Plazaola_Becerril}%
  \BibitemOpen
  \bibfield  {author} {\bibinfo {author} {\bibfnamefont {J.~I.}\ \bibnamefont
  {García‐Plazaola}}\ and\ \bibinfo {author} {\bibfnamefont {J.~M.}\
  \bibnamefont {Becerril}},\ }\bibfield  {title} {\enquote {\bibinfo {title}
  {{A rapid high‐performance liquid chromatography method to measure
  lipophilic antioxidants in stressed plants: simultaneous determination of
  carotenoids and tocopherols}},}\ }\href
  {https://doi.org/10.1002/(sici)1099-1565(199911/12)10:6<307::aid-pca477>3.0.co;2-l}
  {\bibfield  {journal} {\bibinfo  {journal} {Phytochemical Analysis}\ }\textbf
  {\bibinfo {volume} {10}},\ \bibinfo {pages} {307--313} (\bibinfo {year}
  {1999})}\BibitemShut {NoStop}%
\end{thebibliography}%

\end{document}


\setcounter{figure}{0}
\renewcommand{\figurename}{Fig.}
\renewcommand{\thefigure}{S\arabic{figure}}
\setcounter{table}{0}
\renewcommand{\tablename}{Table}
\renewcommand{\thetable}{S\arabic{table}}
\setcounter{equation}{0}
\renewcommand{\theequation}{S\arabic{equation}}

\title{Supporting Information to: ``Kinetics of the Xanthophyll Cycle and its Role in the Photoprotective Memory and Response''}
\author{Audrey Short}
\thanks{These authors contributed equally.}
\affiliation{Graduate Group in Biophysics, University of California, Berkeley, CA 94720 USA}
\affiliation{Molecular Biophysics and Integrated Bioimaging Division Lawrence Berkeley National Laboratory, Berkeley, CA 94720 USA}
\affiliation{Kavli Energy Nanoscience Institute, Berkeley, CA 94720 USA}
\author{Thomas P. Fay}
\thanks{These authors contributed equally.}
\affiliation{Department of Chemistry, University of California Berkeley, CA 94720 USA}
\author{Thien Crisanto}
\affiliation{Molecular Biophysics and Integrated Bioimaging Division Lawrence Berkeley National Laboratory, Berkeley, CA 94720 USA}
\affiliation{Department of Plant and Microbial Biology, University of California, Berkeley, CA 94720 USA }
\affiliation{Howard Hughes Medical Institute, University of California, Berkeley, CA 94720 USA}
\author{Ratul Mangal}
\affiliation{Department of Chemistry, University of California Berkeley, CA 94720 USA}
\author{Krishna K. Niyogi}
\affiliation{Molecular Biophysics and Integrated Bioimaging Division Lawrence Berkeley National Laboratory, Berkeley, CA 94720 USA}
\affiliation{Department of Plant and Microbial Biology, University of California, Berkeley, CA 94720 USA }
\affiliation{Howard Hughes Medical Institute, University of California, Berkeley, CA 94720 USA}
\author{David T. Limmer}
\affiliation{Graduate Group in Biophysics, University of California, Berkeley, CA 94720 USA}
\affiliation{Kavli Energy Nanoscience Institute, Berkeley, CA 94720 USA}
\affiliation{Department of Chemistry, University of California Berkeley, CA 94720 USA}
\affiliation{Chemical Science Division Lawrence Berkeley National Laboratory, Berkeley, CA 94720 USA}
\affiliation{Material Science Division Lawrence Berkeley National Laboratory, Berkeley, CA 94720 USA}
\author{Graham R. Fleming}
\email{grfleming@lbl.gov}
\affiliation{Graduate Group in Biophysics, University of California, Berkeley, CA 94720 USA}
\affiliation{Molecular Biophysics and Integrated Bioimaging Division Lawrence Berkeley National Laboratory, Berkeley, CA 94720 USA}
\affiliation{Kavli Energy Nanoscience Institute, Berkeley, CA 94720 USA}
\affiliation{Department of Chemistry, University of California Berkeley, CA 94720 USA}

\maketitle

\tableofcontents

\section{Further Model details}

The kinetic scheme for our VAZ cycle based model of non-photochemcial quenching in \Nanno\ is given explicitly here
\begin{gather}
    \ce{P + V <=>[$k$_{PV,f}][$k$_{PV,b}] PV <=>[$k_{\mathrm{QV,f}}^\mathrm{light/dark}$][$k_{\mathrm{QV,b}}^\mathrm{light/dark}$] QV  } \\
    \ce{P + A <=>[$k$_{PA,f}][$k$_{PA,b}] PV <=>[$k_{\mathrm{QA,f}}^\mathrm{light/dark}$][$k_{\mathrm{QA,b}}^\mathrm{light/dark}$] QA  } \\
    \ce{P + Z <=>[$k$_{PZ,f}][$k$_{PZ,b}] PV <=>[$k_{\mathrm{QZ,f}}^\mathrm{light/dark}$][$k_{\mathrm{QZ,b}}^\mathrm{light/dark}$] QZ  } \\
    \ce{V +VDEa  <=>[$k$_{V\to A}][$k$_{A\to V}] A +VDEa <=>[$k$_{A\to Z}][$k$_{Z\to A}] Z +VDEa } \\
    \ce{VDEi <=>[$k_{\mathrm{VDE,f}}^\mathrm{light/dark}$][$k_{\mathrm{VDE,b}}^\mathrm{light/dark}$] VDEa }.
\end{gather}
Each step is treated as an elementary rate process in constructing kinetic equations for the chemical species. The full set of kinetic equations is therefore 
\begin{align}
    \dv{t}[\ce{X}] &= -k_{\ce{PX},\mathrm{F,eff}} [\ce{X}][\ce{P}] + k_{\ce{PX},\ce{b}} [\ce{PX}], \ \text{ for X = V,A,Z} \\
    \dv{t}[\ce{PX}] &= k_{\ce{PX},\mathrm{F,eff}} [\ce{X}][\ce{P}] - k_{\ce{PX},\ce{b}} [\ce{PX}] - k_{\ce{QX},\mathrm{F,eff}}^{\mathrm{light/dark}} [\ce{PX}] + k_{\ce{QX},\ce{b}}^{\mathrm{light/dark}}[\ce{QX}], \ \text{ for X = V,A,Z} \\
    \dv{t}[\ce{QX}] &= k_{\ce{QX},\mathrm{F,eff}}^{\mathrm{light/dark}} [\ce{PX}] - k_{\ce{QX},\ce{b}}^{\mathrm{light/dark}}[\ce{QX}], \ \text{ for X = V,A,Z} \\
    \dv{t}[\ce{P}] &= -\sum_{\ce{X} = \mathrm{V,A,Z}}k_{\ce{PX},\mathrm{F,eff}} [\ce{X}][\ce{P}] +\sum_{\ce{X} = \mathrm{V,A,Z}} k_{\ce{PX},\ce{b}} [\ce{PX}] \\
    \dv{t}[\ce{V}] &= - k_{\ce{V}\to\ce{A}} [\ce{VDEa}][\ce{V}] + k_{\ce{A}\to\ce{V}}[\ce{A}] \\
    \dv{t}[\ce{A}] &= k_{\ce{V}\to\ce{A}} [\ce{VDEa}][\ce{V}] - k_{\ce{A}\to\ce{V}}[\ce{A}] - k_{\ce{A}\to\ce{Z}} [\ce{VDEa}][\ce{A}] + k_{\ce{Z}\to\ce{A}}[\ce{Z}] \\
    \dv{t}[\ce{Z}] &=  k_{\ce{A}\to\ce{Z}} [\ce{VDEa}][\ce{A}] - k_{\ce{Z}\to\ce{A}}[\ce{Z}] \\ 
    \dv{t}[\ce{VDEa}] &= - \dv{t}[\ce{VDEi}] = k_{\ce{VDE},\mathrm{F,eff}}^{\mathrm{light/dark}} [\ce{VDEi}] - k_{\ce{VDE},\ce{b}}^{\mathrm{light/dark}} [\ce{VDEa}].    
\end{align}
The light/dark labeled rate constants take different values depending on the light conditions at a time $t$ in a given sequence of HL/D exposures, i.e. 
\begin{align}
    k^{\mathrm{light/dark}} \equiv k^{\mathrm{light/dark}}(t)  = \begin{cases}
    k^{\mathrm{light}}, \text{if HL at time }t \\
    k^{\mathrm{dark}}, \text{if D at time }t.
    \end{cases}
\end{align}
There is some parametric redundancy in fitting the model to \NPQtau\ and HPLC data, specifically the model is independent of scaling $[\ce{VDE}]_{\mathrm{tot}} \to \gamma [\ce{VDE}]_{\mathrm{tot}}$, $k_{\ce{V \to A}} \to (1/\gamma) k_{\ce{V \to A}}$ and $k_{\ce{A \to Z}} \to (1/\gamma) k_{\ce{A \to Z}}$. As such we only work explicitly with the activity of VDE as a dynamical variable,
\begin{align}
    \alpha_{\ce{VDE}}(t) = \frac{[\ce{VDEa}]}{[\ce{VDEa}]_{\mathrm{eq}}^{\mathrm{light}}}, 
\end{align}
where $[\ce{VDEa}]_{\mathrm{eq}}^{\mathrm{light}}$ is the equilibrium concentration of VDEa under light conditions, and we fit the maximum de-epoxidation rates, $k_{\ce{V \to A,max}} = k_{\ce{V \to A}}[\ce{VDEa}]_{\mathrm{eq}}^{\mathrm{light}}$ and $k_{\ce{A \to Z,max}} = k_{\ce{A \to Z}}[\ce{VDEa}]_{\mathrm{eq}}^{\mathrm{light}}$, and the response rate $k_{\mathrm{VDE}}^{\mathrm{light/dark}} = k_{\mathrm{VDE,f}}^{\mathrm{light/dark}} + k_{\mathrm{VDE,b}}^{\mathrm{light/dark}}$. Overall the equation for $\alpha_{\ce{VDE}}(t)$ is
\begin{align}
    \dv{t}\alpha_{\ce{VDE}}(t) = \alpha_{\ce{VDE},\ce{eq}}^{\mathrm{light/dark}} - k_{\mathrm{VDE}}^{\mathrm{light/dark}} (\alpha_{\ce{VDE}}(t)-\alpha_{\ce{VDE},\ce{eq}}^{\mathrm{light/dark}})
\end{align}
where $\alpha_{\ce{VDE},\ce{eq}}^{\mathrm{light}} = 1$ and $\alpha_{\ce{VDE},\ce{eq}}^{\mathrm{dark}} = { [\ce{VDEa}]_{\mathrm{eq}}^{\mathrm{dark}} }/{ [\ce{VDEa}]_{\mathrm{eq}}^{\mathrm{light}} }$.

As stated in the methods section, we work in reduced variables given by $[\widetilde{\ce{B}}] = \tau_{F}(0)k_{\ce{qE}}[\ce{B}]$, where $\tau_{F}(0)$ is the fluorescence lifetime at $t = 0$\, and $k_{\ce{qE}}$ is the quenching rate associated with the QX species. With this the total \NPQtau\ is given by
\begin{align}
    \mathrm{NPQ}_\tau(t) &= \frac{\tau_F(0)-\tau_F(t)}{\tau_F(t)} = \tau_F(0)(\tau_F(t)^{-1} - \tau_F(0)^{-1}) \\
    &= \tau_F(0)k_{\mathrm{qE}}\bigg(\Delta[\ce{QV}](t) + \Delta[\ce{QA}](t) + \Delta[\ce{QZ}](t) + \frac{k_{\mathrm{qZ}}}{k_{\mathrm{qE}}}\Delta[\ce{Z}](t)\bigg) \\
    &=  \Delta[\widetilde{\ce{QV}}](t) + \Delta[\widetilde{\ce{QA}}](t) + \Delta[\widetilde{\ce{QZ}}](t) + \frac{k_{\mathrm{qZ}}}{k_{\mathrm{qE}}}\Delta[\widetilde{\ce{Z}}](t).
\end{align}
In order the model the VDE mutant \NPQtau we account for the fact that the model predicts different fluorescence lifetimes for the WT and \emph{vde} mutant,
\begin{align}
    \mathrm{NPQ}_\tau^{\text{\emph{vde}}}(t) &= \frac{1}{1-[\widetilde{\ce{QV}}]^{\text{\emph{vde}}}(0) + \sum_{\ce{X}}[\widetilde{\ce{QX}}]^{\mathrm{WT}}(0)+({k_{\mathrm{qZ}}}/{k_{\mathrm{qE}}})[\widetilde{\ce{Z}}]^{\mathrm{WT}}(0)} \Delta[\widetilde{\ce{QV}}]^{\text{\emph{vde}}}(t).
\end{align}
We find the correction factor to be almost exactly 1 (1.000006), which agrees with the very similar fluorescence lifetimes of the \emph{vde} and WT species in the initial dark period of the experiments.

In fitting the model parameters we set the rate constants for the P+X binding and unbinding to be independent of the xanthophyll, and we also set the $k_{\ce{QX}}^{\mathrm{light/dark}} = k_{\ce{QX,f}}^{\mathrm{light/dark}} + k_{\ce{QX,b}}^{\mathrm{light/dark}}$ to be the same for all three xanthophylls. This reduces the number of free parameters and ensures that the only parameter controlling the efficacy of the xanthophylls as quencher is the equilibrium constant for the \ce{PX <=> QX} of a given xanthophyll. The parameters treated explicitly as free parameters are those given in Table \ref{tab-params}.

The kinetic equations for the model were solved using the ``ode23s'' solver in Matlab. Model parameters were fit to minimise the least squares difference between the model and experimental \NPQtau\
\begin{align}
    \mathcal{L} = \sum_{s,i}(\mathrm{NPQ}^\mathrm{model}_\tau(t_i;s) - \mathrm{NPQ}^\mathrm{exp}_\tau(t_i;s))^2
\end{align}
where $s$ labels the sequences used in the fitting procedure: the 5 HL-9 D-5 HL, 5 HL-15 D-5 HL, 3 HL-1 D-1 HL-3 D-9 HL-3 D, 1 HL-2 D-7 HL-5 D-1 HL-2 D, 2 HL-2 D sequences. The parameters were fitted first using Matlab's ``global search'' function from an initial guess based on our previous model and HPLC data fits (described below). This was then refined using the ``patternsearch'' algorithm. Errors in the fitted parameters were estimated by bootstrapping the experimental data 1000 times and all reported errors are two standard deviations in the mean of the bootstrapped parameter distributions.

\section{Reduced model for HPLC data}

In order to obtain first estimates of the xanthophyll epoxidation/de-epoxidation rates, we fitted the HPLC data directly to a reduced version of the full. We obtain this reduced model by assuming the binding/unbinding time-scales and \ce{PX <=> QX} time-scales are fast compared to the xanthophyll interconversion. With this we can invoke a quasi-equilibrium approximation for the P,X,PX and QX species.
\begin{align}
    [\ce{QX}] &\approx  K_{\ce{QX}}^{\mathrm{light/dark}} [\ce{PX}] \\
    [\ce{PX}] &\approx  K_{\ce{PX}} [\ce{P}][\ce{X}].
\end{align}
With this we find the pool X concentration is
\begin{align}
    [\ce{X}] &\approx \frac{1}{1+K_{\ce{PX,eff}}[\ce{P}]} [\ce{X}]_{\mathrm{tot}} \\
    K_{\ce{PX,eff}} &= (1+ K_{\ce{QX}}^{\mathrm{light/dark}})K_{\ce{PX}},
\end{align}
and therefore the rate of xanthophyll interconversion is given by
\begin{align}
    \dv{t}[\ce{V}]_{\mathrm{tot}} &= -\alpha_{\ce{VDE}}(t) \frac{k_{\ce{V \to A},\mathrm{max}}}{1+K_{\ce{PV,eff}}[\ce{P}]} [\ce{V}]_{\mathrm{tot}} + \frac{k_{\ce{A \to V}}}{1+K_{\ce{PA,eff}}[\ce{P}]}  [\ce{A}]_{\mathrm{tot}} \\
     \dv{t}[\ce{A}]_{\mathrm{tot}} &= \alpha_{\ce{VDE}}(t) \frac{k_{\ce{V \to A},\mathrm{max}}}{1+K_{\ce{PV,eff}}[\ce{P}]} [\ce{V}]_{\mathrm{tot}} - \frac{k_{\ce{A \to V}}}{1+K_{\ce{PA,eff}}[\ce{P}]}  [\ce{A}]_{\mathrm{tot}} -\alpha_{\ce{VDE}}(t) \frac{k_{\ce{A \to Z},\mathrm{max}}}{1+K_{\ce{PA,eff}}[\ce{P}]} [\ce{A}]_{\mathrm{tot}} + \frac{k_{\ce{Z \to A}}}{1+K_{\ce{PZ,eff}}[\ce{P}]}  [\ce{Z}]_{\mathrm{tot}} \\
    \dv{t}[\ce{Z}]_{\mathrm{tot}} &= \alpha_{\ce{VDE}}(t) \frac{k_{\ce{A \to Z},\mathrm{max}}}{1+K_{\ce{PA,eff}}[\ce{P}]} [\ce{A}]_{\mathrm{tot}} - \frac{k_{\ce{Z \to A}}}{1+K_{\ce{PZ,eff}}[\ce{P}]}  [\ce{Z}]_{\mathrm{tot}}.
\end{align}
Because the ``pool'' xanthophylls are in excess [P] is very small so it can be treated as being in steady state, so we assume that $K_{\ce{PX}}[\ce{P}]$ can be treated as constant. This means estimates of the xanthophyll interconversion rates can be obtained using a first order kinetic model with light-phase dependent rate constants, and the VDE activation as treated in the full model.

\section{Model parameters}

The final set of fitted model parameters are given in Table \ref{tab-params}, obtained from least squares fitting of a subset of the \NPQtau data with xanthophyll interconversion rate constants constrained to be within 50\% of values obtained from the reduced model fitting. The reduced model fitting produced rate constants of $k_{\ce{V \to A,max}} = 0.1307$ min${}^{-1}$, $k_{\ce{A \to Z,max}} = 0.0918$ min${}^{-1}$, $k_{\ce{Z \to A,max}} = 0.1245$ min${}^{-1}$, $k_{\ce{A \to V,max}} = 0.0458$ min${}^{-1}$, $\alpha_{\ce{VDE},\ce{eq}}^\mathrm{dark} = 0.0013$, $k_{\ce{VDE}}^{\mathrm{light}} = 1.285$ min${}^{-1}$, and $k_{\ce{VDE}}^{\mathrm{dark}} = 1.019$ min${}^{-1}$. 

In comparing the model HPLC data to the experimental HPLC data, we found a scaling constant of 0.98 mmol / mol Chl between the reduced units of the model and the concentration relative the Chl by least squares fitting the full model HPLC predictions to the experimental values. From this we can estimate the total concentration of LHCX1 (P in the model) to be about 3.5 mmol / mol Chl. 

\begin{table}[t]
    \centering
    \begin{tabular}{cccc}
    \hline
    Parameter &  Value & Lower bound (95\% CI) & Upper bound (95\% CI)\\
    \hline
k_{\ce{A \to Z,max}}$	& 0.1361 & 0.0935 & 0.1951 \\ 
	$k_{\ce{V \to A,max}}$	& 0.0918 & 0.0688 & 0.1181 \\ 
	$k_{\ce{Z \to A}}$	& 0.0854 & 0.0832 & 0.1414 \\ 
	$k_{\ce{A \to V}}$	& 0.0509 & 0.0307 & 0.0685 \\ 
	$k_{\ce{VDE}}^{\mathrm{light}}$	& 1.2846 & 1.1954 & 1.2962 \\ 
	$k_{\ce{VDE}}^{\mathrm{dark}}$	& 1.0193 & 0.5732 & 76.4352 \\ 
	$\alpha_{\ce{VDE,eq}}^{\mathrm{dark}}$	& 0.0010 & 0.0009 & 0.0019 \\ 
	$k_{\ce{PV,b}}$	& 3.4187 & 1.6648 & 10.8108 \\ 
	$k_{\ce{PA,b}}$	& 3.4187 & 1.6648 & 10.8108 \\ 
	$k_{\ce{PZ,b}}$	& 3.4187 & 1.6648 & 10.8108 \\ 
	$k_{\ce{QZ}}^{\mathrm{light}}$	& 2.0744 & 1.9304 & 2.3957 \\ 
	$k_{\ce{QZ}}^{\mathrm{dark}}$	& 4.6913 & 4.3656 & 8.3424 \\ 
	$K_{\ce{PV}}$	& 0.2400 & 0.2079 & 1623.4242 \\ 
	$K_{\ce{PA}}$	& 0.2400 & 0.2079 & 1623.4242 \\ 
	$K_{\ce{PZ}}$	& 0.2400 & 0.2079 & 1623.4242 \\ 
	$K_{\ce{QZ}}^{\mathrm{light}}$	& 10.9158 & 3.4519 & 14.5586 \\ 
	$K_{\ce{QZ}}^{\mathrm{dark}}$ &	0 & -- & -- \\ 
	$[\widetilde{\ce{V}}]_0$	& 67.9332 & 67.6284 & 69.1663 \\ 
	$[\widetilde{\ce{P}}]_\mathrm{tot}$	& 3.5324 & 3.5245 & 3.5326 \\ 
	$K_{\ce{QA}}^{\mathrm{light}}$	& 0.3872 & 0.0353 & 0.5163 \\ 
	$K_{\ce{QA}}^{\mathrm{dark}}$	&  0 & -- & -- \\ 
	$k_{\ce{QA}}^{\mathrm{light}}$	& 2.0744 & 1.9304 & 2.3957 \\ 
	$k_{\ce{QA}}^{\mathrm{dark}}$	& 4.6913 & 4.3656 & 8.3424 \\ 
	$K_{\ce{QV}}^{\mathrm{light}}$	& 0.1173 & 0.0879 & 0.1175 \\ 
	$K_{\ce{QV}}^{\mathrm{dark}}$ &	0 & -- & -- \\ 
	$k_{\ce{QV}}^{\mathrm{light}}$	& 2.0744 & 1.9304 & 2.3957 \\ 
	$k_{\ce{QV}}^{\mathrm{dark}}$	& 4.6913 & 4.3656 & 8.3424 \\ 
 $k_{\ce{qZ}}/k_{\mathrm{qE}}$ & 0.0259 & 0.0234 & 0.0343 \\
 \hline
    \end{tabular}
    \caption{Best fit parameters obtained for the full model. Confidence intervals obtained by bootstrapping experimental NPQ runs and estimating 95\% confidence intervals from the approximate parameter distribution. All parameters are given in reduced units of the model, therefore all rate constants are in min${}^{-1}$ and all other parameters are unitless.}
    \label{tab-params}
\end{table}



\section{Mechanism of qZ}

In our model we treat the qZ quenching process as an additional first order quenching process just proportional to the concentration of ``pool'' Zeaxanthin. We can arrive at this model using a simple model similar to our LHCX1 based quenching model. We consider adding a second protein or complex to our model denoted \ce{P$'$}, which binds xanthophylls to form complexes \ce{PX$'$} = \ce{PV$'$}, \ce{PA$'$}, \ce{PZ$'$}. We assume the quenching of chlorophyll excitations is proportional to the concentration of \ce{PZ$'$}, such that the change in fluroescence decay rate is $\Delta k_{\mathrm{F,qZ}} = k_{\mathrm{Q,PZ'}} [\ce{PZ}']$. Assuming that P' binding X can be treated witht he pre-equilibrium/quasi-equilibrium approximation, we find that
\begin{align}
    [\ce{PX}'] = \frac{K_{\ce{PX}'[\ce{P'}]}[\ce{X}]_\mathrm{pool}}{K_{\ce{PX}'}[\ce{P'}] + 1}.
\end{align}
where $[\ce{X}]_\mathrm{pool}$ is the xanthophyll concentration in the pool including that bound to P', and $K_{\ce{PX}}$ is the equilibrium constant for P' binding X. Assuming that P' is in a steady state, where $\dv{t}[\ce{P'}](t) \approx 0$, and thus $[\ce{P'}](t) \approx [\ce{P'}]_0$, the change in quenching rate due to qZ is simply proportional to $[Z]_{\mathrm{pool}}$, as is assumed in the model.

\section{Estimating quenching rates}

We can construct a simple model for excitation quenching as follows. We assume that the excited chlorophylls, \ce{Chl^*}, can exist either on an active quenching complex, QX, which we label \ce{Chl^*_{Q}}, or on the other light-harvesting complexes, which we label \ce{Chl^*_{Pool}}. We treat the populations of \ce{Chl^*} in these two environment with a simple first order kinetic model, with a diffusion rate onto QX of $\eta_{\ce{Q}} k_{\ce{D}}$ and a diffusion rate off the QX site given by $k_{\ce{D}}$. $\eta_{\ce{Q}}$ is the ratio of the number of Chl on QX to the number of Chl in the whole system, which we estimate to be approximately the ratio of QX to the all of the light harvesting proteins. We further assume that the rate of decay of the \ce{Chl^*} down to its ground state is dependent on the site, occurring at a rate $k_{\mathrm{F,eff},0}$ (\ce{Chl^*} decay is dominated by non-radiative decay, but this rate constant should be understood as including a small radiative contribution) in the pool and at a rate $k_{\mathrm{F,eff},\ce{Q}}$ on the quenching sites. Putting these ingredient together we arrive at the following kinetic equations for \ce{Chl^*_{Q}} and \ce{Chl^*_{Pool}}
\begin{align}
    \dv{t}[\ce{Chl^*_{Q}}] &= -(k_{\mathrm{F,eff},\ce{Q}} + k_{\ce{D}}) [\ce{Chl^*_{Q}}] + \eta_{\ce{Q}}k_{\ce{D}} [\ce{Chl^*_{Pool}}] \\
    \dv{t}[\ce{Chl^*_{Pool}}] &= -(k_{\mathrm{F,eff},0} + \eta_{\ce{Q}} k_{\ce{D}}) [\ce{Chl^*_{Pool}}] + k_{\ce{D}} [\ce{Chl^*_{Q}}].
\end{align}
Applying the steady state approximation to $[\ce{Chl^*_{Q}}]$, we obtain the following equation for the decay of the pool \ce{Chl^*},
\begin{align}
    \dv{t}[\ce{Chl^*_{Pool}}] &\approx - \left(k_{\mathrm{F,eff},0} +  \eta_{\ce{Q}} k_{\ce{D}} \frac{k_{\mathrm{F,eff},\ce{Q}}}{k_{\mathrm{F,eff},\ce{Q}}+k_{\ce{D}}} \right) [\ce{Chl^*_{Pool}}],
\end{align}
from which we obtain the fluorescence lifetime as 
\begin{align}
    \frac{1}{\tau_{\mathrm{F,eff}}} = k_{\mathrm{F,eff},0} +  \eta_{\ce{Q}} k_{\ce{D}} \frac{k_{\mathrm{F,eff},\ce{Q}}}{k_{\mathrm{F,eff},\ce{Q}}+k_{\ce{D}}},
\end{align}
recalling that $\eta_{\ce{Q}} \propto \sum_{\ce{X}}[\ce{QX}]$, the expression we find is consistent with the assumptions of our NPQ model (excluding qZ). Assuming $\eta_{\ce{Q}} \approx 0$ before light exposure, we find the \NPQtau\ as
\begin{align}
    \mathrm{NPQ}_\tau = \eta_{\ce{Q}}\frac{k_{\ce{D}}}{k_{\mathrm{F,eff},0}}\frac{k_{\mathrm{F,eff},\ce{Q}}}{k_{\mathrm{F,eff},\ce{Q}}+k_{\ce{D}}}.
\end{align}
If we assume excitation energy diffusion is very fast between proteins compared to the other time-scales in the model, we find that the \NPQtau] is given approximately by 
\begin{align}
    \mathrm{NPQ}_\tau = \eta_{\ce{Q}}\frac{k_{\mathrm{F,eff},\ce{Q}}}{k_{\mathrm{F,eff},0}}.
\end{align}
From the time-correlated photon counting experiments used to obtain the \NPQtau\ we know $k_{\mathrm{F,eff},0} \approx 1 \text{ ns}^{-1}$. The maximum \NPQtau\ within our model is limited by the total concentration of P (in reduced units), $[\widetilde{\ce{P}}]_\mathrm{tot}\sim\! 3.5$. From the HPLC experiments we have deduced that P is present at a concentration of around 3.5 mmol / mol Chl. Assuming $\sim\! 10$ Chl per light-harvesting protein, this means about 1 in 30 proteins in the chloroplast would be P, which puts an upper bound on $\eta_{\ce{Q}}$ of $\sim 1/30$. From this we can estimate a lower bound on $k_{\mathrm{F,eff},\ce{Q}}$ to be $k_{\mathrm{F,eff},\ce{Q}} \sim 100 \text{ ns}^{-1}$, i.e. the lifetime of \ce{Chl^*} on the quencher must be $\sim 10 $ ps. If we instead use (1/7.7) ps${}^{-1}$ as an estimate for $k_{\mathrm{F,eff},\ce{Q}}$, as obtained in Ref.~\onlinecite{park2019a}, we deduce that roughly 1 in 43 light-harvesting proteins in the chloroplast are P. Given the large simplifications and the uncertainty in the abundance of P deduced from HPLC data and the model (due to the large uncertainty in the conversion factor from model concentration to abundance in the thylakoid membrane), we consider these estimates of the proportion of P and the quenching lifetimes as being in excellent agreement.

\section{Raw HPLC data}

\begin{figure}[t]
    \centering
    \includegraphics[width=0.8\textwidth]{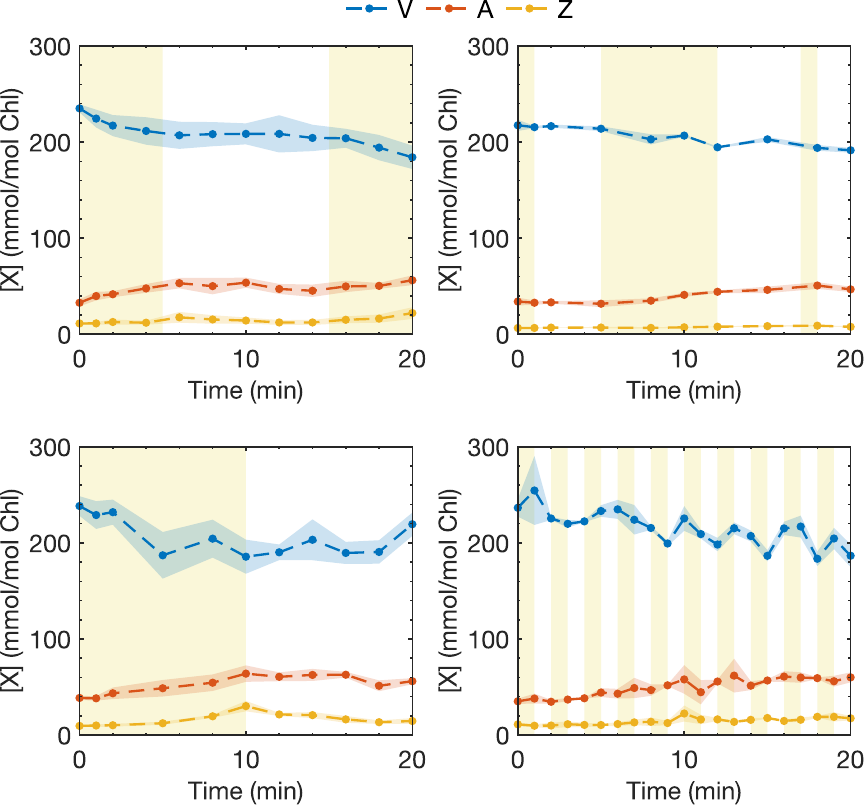}
    \caption{Raw HPLC data for concentrations of each xanthophyll normalized by the total Chl concentration for four HL/D sequences: 5 HL- 10 D- 5 HL (top left), 1 HL- 4 D- 7 HL- 5 D- 1 HL- 2 D (top right), 10 HL- 10 D (bottom left), 1 HL-1 D (bottom right).}
    \label{fig-hplc-raw}
\end{figure}

In Fig.~\ref{fig-hplc-raw} we show the raw HPLC data for each of the HL/D sequences shown in the main text. A certain fraction of each xanthophyll does not change over the course of the experiment. Since our model only includes xanthophylls that are free to bind/unbind from proteins on the time-scale of our experiments, we only examine the changes in xanthophyll concentration, and use these changes in fitting the model.

\section{NPQ recovery in 5 HL-$T$ D-5 HL sequences}

\begin{figure}
    \centering
    \includegraphics[width=0.65\textwidth]{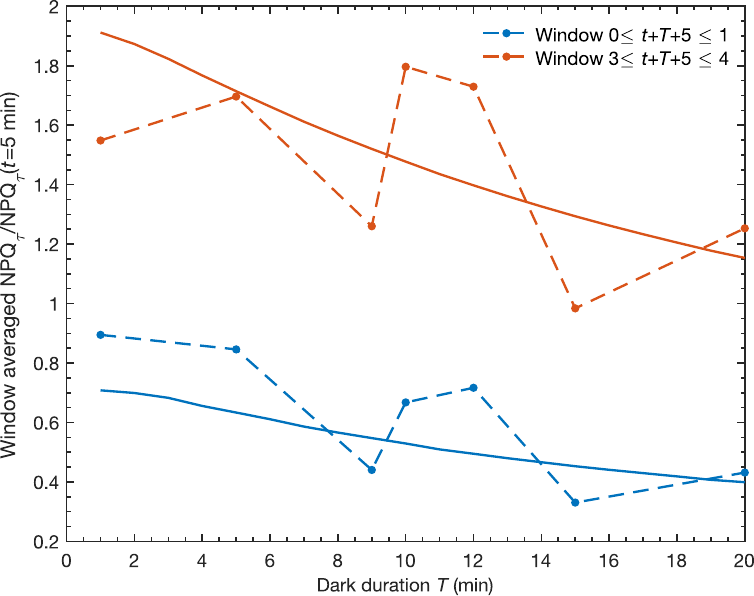}
    \caption{Window averaged \NPQtau\ in the second light phase for the 5 HL-$T$ D-5 HL sequences, normalised by the \NPQtau\ value at $t = 5$ min, for the first minute (blue) and fourth minute (red) of the second light phase, from the experiment (dashed lines and circles) and model (solid line).}
    \label{fig-npq-rec-av}
\end{figure}

In Fig.~\ref{fig-npq-rec-av} we show window averaged \NPQtau\ in the second light phase for the 5 HL-$T$ D-5 HL sequences, normalised by the \NPQtau\ value at $t = 5$ min. This window averaging is defined as
\begin{align}
    \overline{\mathrm{NPQ}}_\tau = \frac{1}{t_f-t_i}\int_{t_i}^{t_f} \mathrm{NPQ}_\tau(t)\dd{t}.
\end{align}
The experimental window averaging is estimated using the trapezoidal rule. 
Fitting the averaged normalised \NPQtau\ in the first minute to an exponential decay as a function of $T$, i.e. $\overline{\mathrm{NPQ}_\tau} = \overline{\mathrm{NPQ}_\tau,0} e^{-k_\mathrm{mem} T}$, we obtain an effective recovery rate constant of $k_\mathrm{mem} = 0.0464$ (lower CI (95\%): $0.0064$, upper CI (95\%): $0.0861$) min${}^{-1}$, which matches the model $k_{\ce{A -> V}}$ rate constant of 0.0509 min${}^{-1}$.

\section*{References}

\bibliography{si/references-si.bib}